\documentclass[12pt,a4paper]{article}
\usepackage[T1]{fontenc}
\usepackage[english]{babel}     %LRY: option 'russian' removed
\usepackage{graphicx}
\usepackage{amssymb}
\usepackage{amsmath}
\usepackage{caption}

\DeclareCaptionLabelFormat{viadot}{#1 #2.}            %LRY: do not touch 
\usepackage{subcaption}
\usepackage{indentfirst}
\usepackage[utf8]{inputenc} 
\usepackage{cite}
\usepackage{fullpage}
\usepackage{authblk}
\usepackage[colorlinks=true,linkcolor={blue},citecolor={green},urlcolor={red}]{hyperref}

\renewcommand{\vec}[1]{{\boldsymbol{#1}}}

\newcommand{\fr}[2]{{\displaystyle \frac{#1}{#2}}}
\newcommand{\sfr}[2]{{{#1}/{#2}}}

\newcommand{\pdiff}[2]{{\fr{\partial{#1}}{\partial{#2}}}}

\newcommand{\spdiff}[2]{{\sfr{\partial{#1}}{\partial{#2}}}}

\title{\textbf{On possible types of magnetospheres of hot Jupiters}}

\author{%
Zhilkin, A.G.\thanks{E-mail: zhilkin@inasan.ru}, %
Bisikalo D.V.\\%
\textit{\small Institute of Astronomy of Russian Academy of Sciences, Moscow, Russia}\\%
}%
\date{}

\begin{document}
%%%%%%%%%%%%%%%%%                     LRY:    added, do not touch
\renewcommand{\figurename}{Fig.}
\renewcommand{\tablename}{Table}
%\renewcommand{\captionlabeldelim}{.~}
%%%%%%%%%%%%%%%%%%%%%%%%%%%%%%%%%%%%%%%%%%%%%%%%

\maketitle

%\begin{abstract}            
\noindent
{\bf ABSTRACT.} We show that the orbits of exoplanets of the ``hot Jupiter'' type, as a rule,
are located close to the Alf\'{v}en point of the stellar wind of the parent
star. At this, many hot Jupiters can be located in the sub-Alf\'{v}en zone in
which the magnetic pressure of the stellar wind exceeds its dynamic pressure.
Therefore, magnetic field of the wind must play an extremely important
role for the flow of the stellar wind around the atmospheres
of the hot Jupiters. This factor must be considered both in theoretical models and in the
interpretation of observational data. The analysis shows that many typical hot
Jupiters should have shock-less intrinsic
magnetospheres, which, apparently, do not
have counterparts in the Solar System. Such magnetospheres are characterized,
primarily, by the absence of the bow shock, while the magnetic barrier
(ionopause) is formed by the induced currents in the upper layers of the ionosphere.
We confirmed this inference by the three-dimensional numerical simulation of the flow
of the parent star stellar wind around the hot Jupiter HD 209458b in which we took into
account both proper magnetic field  of the planet and magnetic field of the wind.
%\end{abstract}

\section{Introduction}

When celestial bodies that have their own magnetic field interact with
surrounding ionized matter, they create a cavity around themselves, which is
named \emph{magnetosphere}. In particular, such magnetospheres have Solar
System planets blown over by the solar wind plasma \cite{Belenkaya2009}. The
magnetosphere can have a complex structure and to change over the time, due to the
heterogeneity and nonstationarity of the solar wind. Magnetic field of the
planet prevents direct penetration of the solar wind plasma into the atmosphere.
The boundary of the magnetosphere is a relatively thin current layer
(\emph{magnetopause}), which separates proper magnetic field of the planet from
the magnetic field of the solar wind. Location of the magnetopause is determined
by the balance of total pressure (the sum of the dynamic, gas, and magnetic
pressure) from both sides of the boundary. However, in most cases, the total
pressure from the outer side is equal to the dynamic pressure, while the pressure
from the side of the planet is equal to the magnetic one. Such a situation exists,
for instance, in the case of the Earth’s magnetosphere \cite{smhd}. 
In the front of the magnetopause \emph{a bow shock} forms, due to the 
supersonic flow regime. Between shock and magnetopause,
a transition region exists, in which the plasma of the wind is heated, compressed,
and retarded, changing the direction of the motion.  
An extended \emph{magnetospheric tail} forms at the night side.

However, thousands of currently known exoplanets must have magnetospheres too.
Magnetospheres of the exoplanets may have their own specific features. In the
present paper, we will focus on the structure of the magnetospheres of the hot
Jupiters. ``Hot Jupiters'' are exoplanets with the mass comparable to the mass
of Jupiter, located in the immediate vicinity of the parent star \cite{Murray2009}. The
first hot Jupiter was discovered in 1995 \cite{Mayor1995}. Because of the
proximity of the planets to the parent stars and their relatively large size, 
gas envelopes of the hot
Jupiters can overflow their Roche lobes, leading to the formation of the outflows
from the vicinities of the Lagrange points $L_1$ and $L_2$ \cite {Lai2010, Li2010}.
This is indirectly indicated by excessive absorption in the near-UV 
range, observed for some planets \cite{Vidal2003, Vidal2008, BenJaffel2007,
Vidal2004, BenJaffel2010, Linsky2010}. These conclusions are confirmed
theoretically by the one-dimensional aeronomic models \cite{Murray2009, Yelle2004,
Garcia2007, Koskinen2013, Ionov2017}.

Based on the three-dimensional numerical modeling, it was shown in the series of
studies \cite{Bisikalo2013a, Bisikalo2013, Cherenkov2014, Bisikalo2016,
Cherenkov2017, Cherenkov2018, Bisikalo2018} that, depending on the parameters of
a hot Jupiter, gas envelopes of three main types can form \cite{Bisikalo2013}.
To the first type belong \emph{closed envelopes}, in which planet atmosphere
resides inside its Roche lobe. Into the second type fit \emph{open envelopes}
that are formed by the outflows from the nearest Lagrange points. Finally, one
can distinguish \emph{quasi-closed envelopes} of an intermediate type, for which
stellar wind dynamic pressure stops the outflow outside the Roche lobe. Calculations
have shown that in the cases of closed and quasi-closed envelopes the rate of
mass loss by hot Jupiters is significantly lower, compared to the case of open
envelopes. Arakcheev et al. \cite{Arakcheev2017} presented results of numerical
modeling of the flow structure in the vicinity of the hot Jupiter WASP~12b that
took into account the influence of the planet’s magnetic field.
It was shown that the presence of even a relatively weak planet's magnetic field
(the magnetic moment comprised 10 per cent of Jupiter’s magnetic moment)
may lead to a noticeable decrease of the rate of mass loss,  compared to the net
gas-dynamical case. In addition,  magnetic field may cause fluctuations
in the outer parts of the envelope \cite{Bisikalo2017}.

There is an unaccounted factor in the studies quoted above, related to
the magnetic field of the stellar wind. However, the analysis performed in 
the present
paper showed that it is very important. The fact is that, apparently, many hot
Jupiters are located in the sub-Alf\'{v}en zone of the stellar wind, where the
magnetic pressure exceeds the dynamic one. Therefore, accounting for
magnetic field of the wind, formally, switches the flow of the wind around a hot
Jupiter from supersonic regime to subsonic one. As a result, in this mode, bow
shock should not form in the front of the atmosphere \cite{Ip2004}, i.e., the
flow is shock-less. This conclusion follows from the assumption that the magnetic
field of the wind is determined by the average magnetic field at the surface of
the Sun, which is about 1~G. However, magnetic fields of solar type stars can
range from about 0.1 to several G \cite{Fabbian2017, Lammer2012}. In addition,
hot Jupiters can have parent stars not of solar type only, because their
spectral types are from F to M. The azimuthal component of the magnetic field of
the stellar wind is determined by the angular velocity of the stellar spin, which, in
turn, also depends on the spectral type \cite{Lammer2012}. If all these
additional factors are taken into account, it appears that some hot Jupiters may
be located not only in the transition region, separating the sub-Alf\'{v}en and
super-Alf\'{v}en zones, but to move even into the super-Alf\'{v}en zone. This
circumstance considerably expands the set of possible configurations of the
magnetospheres of hot Jupiters.

It should be noted that there is a simple way to approximate account of magnetic
wind field in the net gas-dynamical calculations. To do this, instead of gas pressure
$P$,  one needs to use  full pressure $P_\text{T} = P + B^2/(8\pi)$. It is not difficult to
see that this is equivalent to the replacement of the temperature $T$ of the wind by  
the temperature
\begin{equation}\label{eq-int1}
 \tilde{T} = T \left( 1 + \frac{u_A^2}{2 c_T^2} \right),
\end{equation}
where $c_T$ is the isothermal sound velocity, $u_A$ is the Alf\'{v}en velocity. 
In this case, the spatial distribution of the magnetic field $B$ and, consequently, of the 
temperature $\tilde{T}$ should be determined by some kind of magnetohydrodynamical
wind model. Such an amendment can effectively increase the wind temperature and
to switch the flow around the planet into subsonic mode.

In the present paper, we analyzed the possible types of magnetospheres of hot Jupiters,
taking into account possible outflows resulting from Roche lobe overflow.
The results of numerical simulations using three-dimensional
magnetohydrodynamical model confirm the conclusions obtained on the basis of
simple theoretical considerations.

The structure of the article is as follows. In \S2 we describe
the model of magnetic field of the stellar wind applied by us. In \S3 we
analyze possible types of magnetospheres of hot Jupiters. In \S4 
numerical model is described. In \S5 results of 
calculations are presented. The summary of the main results of the study follows in \S6. 

\section{The model of  magnetic field of the stellar wind}

\begin{figure}
\centering
\includegraphics[width = 0.45\textwidth]{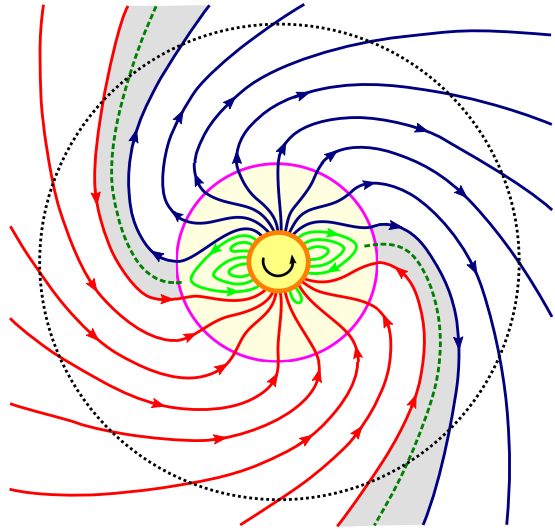}
\caption{Schematic picture of the solar wind in the ecliptic plane. Small dyed
circle in the center corresponds to the Sun. The arrow shows the direction of
rotation of the Sun. The boundary of the middle circle defines the corona
region, at the border of which magnetic field becomes completely radial. The
shaded gray areas correspond to the zones of the heliospheric current sheet
which is shown by dashed lines running from the corona to the periphery. It
separates magnetic field of the solar wind with different directions of magnetic
field lines (from the Sun or to the Sun). The orbit of the planet (shown by a
dotted circle) is located in the heliospheric region. (A color version of the
Figure is available in the electronic version of the journal.}%
\label{fg1}
\end{figure}

In our numerical model, we will rely on the well-studied properties of the solar
wind. As it is shown by numerous ground- and space-based studies (see, for
instance, the recent review \cite{Owens2013}), magnetic field of the solar wind
has a rather complex structure. Schematically, this structure is shown in
Fig.~\ref{fg1}. In the corona region, magnetic field is essentially
non-radial, because there it is mainly defined by the own magnetic
field of the Sun. At the border of the corona, at the distance of several 
solar radii, the field, to a large accuracy, becomes completely radial. Farther, 
\emph {heliospheric region} is located, in which the magnetic field to a substantial
extent is determined by the properties of the solar wind. In the heliosphere,
with the distance from the center,  magnetic field lines gradually twist into a spiral
due to the rotation of the Sun and therefore (especially at large distances) the magnetic
wind field can be described with a good accuracy using the simple Parker model \cite{Parker1958}.

However, the observed magnetic field in the solar wind is not axisymmetric and
has a pronounced sector structure. This is due to the fact that in the different
points of the spherical surface of the corona the field may have different
polarity (direction of the magnetic field lines relative to the direction of the
normal vector), for example, due to the inclination of the solar magnetic axis
to the axis of its rotation. As a result, in the ecliptic plane, in the solar
wind form two clearly distinguished sectors with different magnetic field
directions. In one sector, magnetic field lines are directed to the Sun, while
in the opposite sector they are directed from the Sun. These two sectors are
separated by \emph{heliospheric current sheet}, which is shown in Fig.~\ref{fg1}
in gray. To the current sheet itself correspond two twisted dashed lines going
from the border of the corona to the periphery of the heliosphere. The
heliospheric current sheet rotates with the Sun and, therefore, the Earth, as it
moves around the Sun, crosses it many times a year (about 13 times), moving from the solar wind
sector with one polarity of the magnetic field into the adjacent sector with
opposite polarity of the field.

In this paper we do not take into account possible sectoral structure of the
wind magnetic field, focusing on the impact of its global parameters. More
detailed consideration and account of the unquestionably important effects
associated with the transition of the planets through the current sheet and the
polarity reversal of the magnetic field, we plan to carry out in the further
studies. On the top of everything else, in our model we will assume that the orbit
of a hot Jupiter is located in the heliospheric region beyond the border of the
corona. In Fig.~\ref {fg1} the orbit of the planet is shown as a large dotted
circle.

To the first approximation, to describe the magnetic field $\vec{B}$ 
of the wind in the heliospheric region, 
one can use the simple axisymmetric model described by
Baranov and Krasnobayev in the monograph \cite{Baranov1977}. In the inertial reference frame
 in the spherical coordinates ($r$, $\theta$, $\varphi$) magnetic field and
stellar wind velocity can be represented as follows:
\begin{equation}\label{eq-wmf1}
 \vec{B} = B_r(r) \vec{n}_r + B_{\varphi}(r, \theta) \vec{n}_{\varphi}, \quad
 \vec{v} = v_r \vec{n}_r + v_{\varphi}(r, \theta) \vec{n}_{\varphi}.
\end{equation}
At difference to \cite{Baranov1977}, in these expressions we took into account the dependence of 
$B_{\varphi}$ and $v_{\varphi}$ on the angle  $\theta$, since our model is
three-dimensional. More, for simplicity, in the vicinity of the planet we will consider the radial
component of velocity as constant and equal to $v_w$.

In such an approximation, the structure of stellar wind is described by a system of equations 
containing equation of continuity
\begin{equation}\label{eq-wmf2}
 \frac{1}{r^2} \pdiff{}{r} \left( r^2 \rho v_r \right) = 0,
\end{equation}
Maxwell equation  ($\nabla \cdot \vec{B} = 0$)
\begin{equation}\label{eq-wmf3}
 \frac{1}{r^2} \pdiff{}{r} \left( r^2 B_r \right) = 0,
\end{equation}
equation for angular momentum
\begin{equation}\label{eq-wmf4}
 \frac{\rho v_r}{r} \pdiff{}{r} \left( r v_{\varphi} \right) = 
 \frac{B_r}{4\pi r} \pdiff{}{r} \left( r B_{\varphi} \right)
\end{equation}
and equation of induction
\begin{equation}\label{eq-wmf5}
 \frac{1}{r} \pdiff{}{r} \left( 
 r v_r B_{\varphi} - r v_{\varphi} B_r 
 \right) = 0.
\end{equation}

From the equation of continuity  \eqref{eq-wmf2} we find
\begin{equation}\label{eq-wmf6}
 \rho = \rho_w \left( \frac{A}{r} \right)^2,
\end{equation}
where $A$ is the large semiaxis of the orbit of the planet, 
$\rho_w$ --- the density of the stellar wind at the orbit of the planet.
From the Maxwell equation \eqref{eq-wmf3} we obtain
\begin{equation}\label{eq-wmf7}
 B_r = B_0 \left( \frac{R_s}{r} \right)^2 = B_w \left( \frac{A}{r} \right)^2,
\end{equation}
where $R_s$ is the radius of the star, 
$B_0$ --- field strength at the stellar surface, 
$B_w$ --- radial component of the field at the orbit of the planet.

Let note that from the Eqs.~\eqref{eq-wmf1} and \eqref{eq-wmf2} it follows
\begin{equation}\label{eq-wmf8}
 \frac{B_r}{4\pi\rho v_r} = \frac{r^2 B_r}{4\pi r^2 \rho v_r} = \text{const}.
\end{equation}
This circumstance allows to obtain from the Eqs.~\eqref{eq-wmf3} and \eqref{eq-wmf4} 
two integrals of motion:
\begin{equation}\label{eq-wmf9}
 r v_{\varphi} - \frac{B_r}{4\pi\rho v_r} r B_{\varphi} = L(\theta),
\end{equation}
\begin{equation}\label{eq-wmf10}
 r v_r B_{\varphi} - r v_{\varphi} B_r = F(\theta).
\end{equation}
The function $F(\theta)$ may be found from the boundary conditions at the stellar surface 
($r = R_s$):
\begin{equation}\label{eq-wmf11}
 B_{\varphi} = 0, \quad 
 B_r = B_0, \quad
 v_{\varphi} = \Omega_s R_s \sin\theta,
\end{equation}
where  $\Omega_s$ is angular velocity of the stellar spin. Therefore, 
\begin{equation}\label{eq-wmf12}
 F(\theta) = -\Omega_s R_s^2 \sin\theta B_0 = -\Omega_s r^2 \sin\theta B_r.
\end{equation}

Taking into account the latter expression, solutions of Eqs.~\eqref{eq-wmf3} and \eqref{eq-wmf4} 
may be written as:
\begin{equation}\label{eq-wmf13}
 v_{\varphi} = 
 \frac{\Omega_s \sin\theta r - \lambda^2 L(\theta)/r}{1 - \lambda^2},
\end{equation}
\begin{equation}\label{eq-wmf14}
 B_{\varphi} = 
 \frac{B_r}{v_r} \lambda^2 
 \frac{\Omega_s \sin\theta r - L(\theta)/r}{1 - \lambda^2}.
\end{equation}
Here, $\lambda$ is Alf\'{v}en Mach number for the radial components of the velocity and magnetic field, 
\begin{equation}\label{eq-wmf15}
 \lambda^2 = \frac{4\pi\rho v_r^2}{B_r^2}.
\end{equation}
Close to the stellar surface, the radial wind velocity $v_r$ should be lower than the 
Alf\'{v}en velocity $u_A = |B_r|/\sqrt{4\pi\rho}$\ and the parameter $\lambda <1 $.
At large distances, the radial velocity $v_r$, on the contrary, exceeds
Alf\'{v}en velocity $u_A$ ($\lambda > 1$). This means that at certain
distance from the center of the star $r=a$ (Alf\'{v}en point) the parameter
$\lambda$=1. The domain $r < a$ can be called
\emph{sub-Alf\'{v}en} zone of the stellar wind, and the region $r > a$ ---
\emph{super-Alf\'{v}en} zone, respectively.

The values $v_{\varphi}$ and $B_{\varphi}$ in the expressions \eqref{eq-wmf13} and 
\eqref{eq-wmf14} should be continuous in the Alf\'{v}en point $r = a$. 
Therefore, it is necessary to set 
\begin{equation}\label{eq-wmf16}
 L(\theta) = \Omega_s \sin\theta a^2.
\end{equation}
As a result, we find the final solution 
\begin{equation}\label{eq-wmf17}
 v_{\varphi} = 
 \Omega_s \sin\theta r\,
 \frac{1 - \lambda^2 a^2 / r^2}{1 - \lambda^2},
\end{equation}
\begin{equation}\label{eq-wmf18}
 B_{\varphi} = 
 \frac{B_r}{v_r} 
 \Omega_s \sin\theta r 
 \lambda^2\, 
 \frac{1 - a^2 / r^2}{1 - \lambda^2}.
\end{equation}
We use these relations in the numerical model to describe the stellar wind. 

\begin{figure}
\centering
\includegraphics[width = 0.49\textwidth]{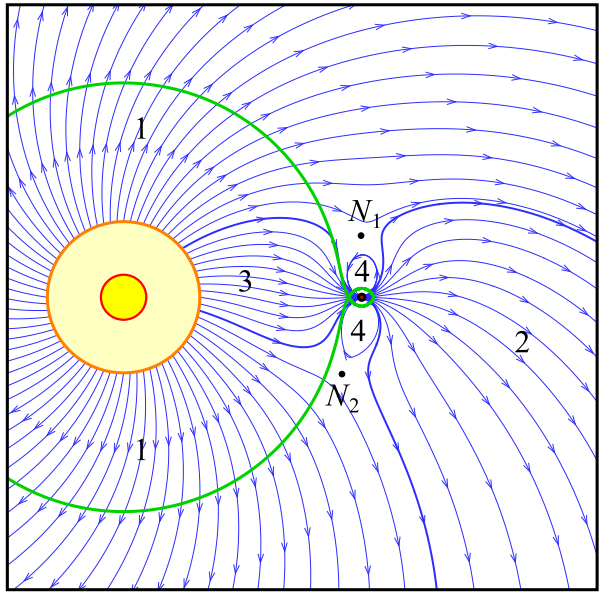}
\caption{Initial distribution of the magnetic field in the equatorial plane for
the case $B_0 = 10^{-3}$~G. Thick solid line shows the Roche lobe. The star is
indicated by a color ring, the inner radius of which corresponds to the radius
of the star and the outer radius --- to the radius of the corona. The numbers in
the diagram label four magnetic zones. Neutral points are marked as
$N_1$ and $N_2$. (A color version of the Figure is available in the electronic
version of the journal.)}%
\label{fg2a}
\end{figure}

\begin{figure}
\centering
\includegraphics[width = 0.49\textwidth]{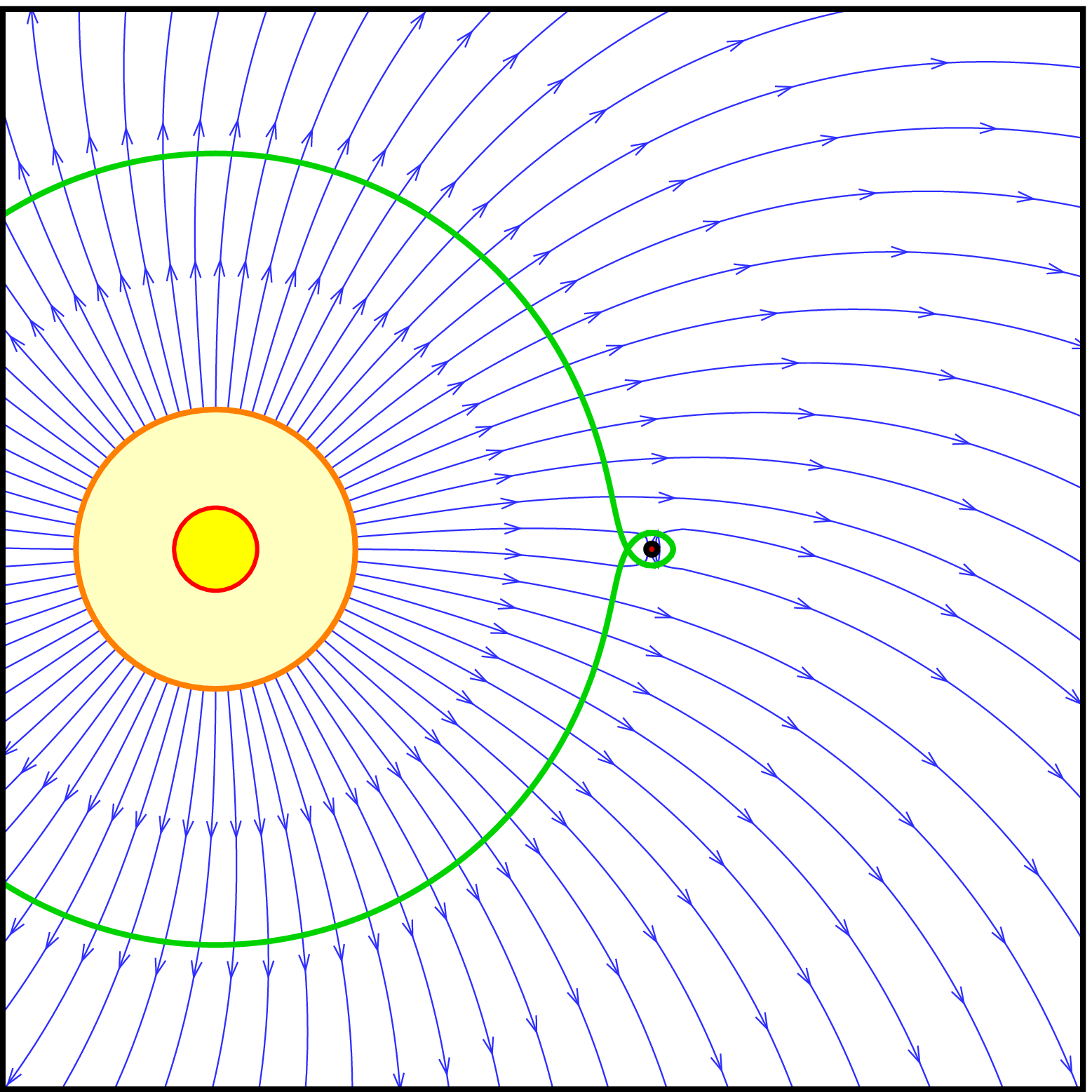}
\includegraphics[width = 0.49\textwidth]{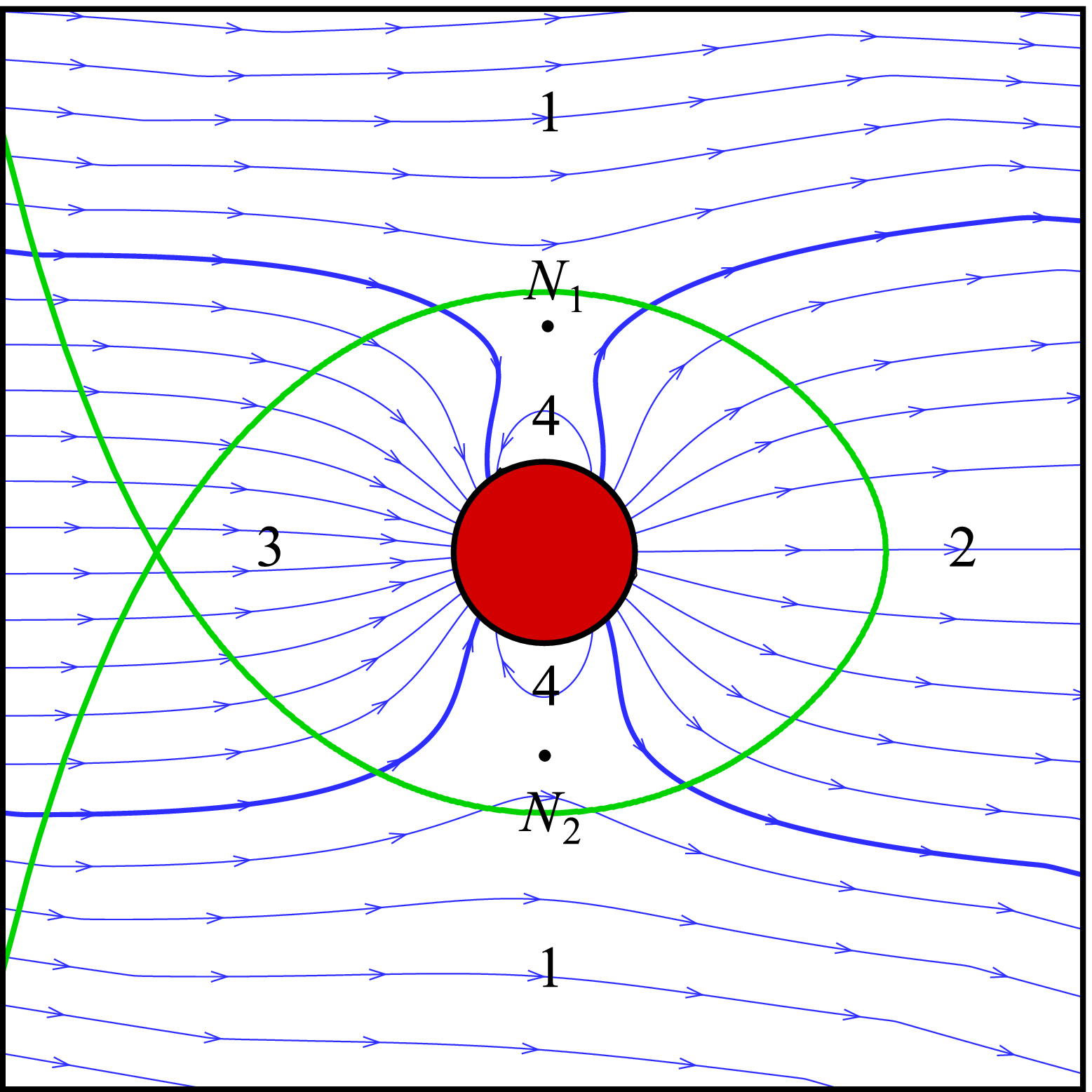}
\caption{Initial distribution of the magnetic field in the equatorial plane for
the case $B_0 = 1$~G. Vicinity of the planet is shown in the blown-up image 
to the right. Notation as in Fig.~\ref{fg2a}.}%
\label{fg2b}
\end{figure}

In Figs. \ref{fg2a} and \ref{fg2b} we show initial (without the outflows from
the envelope) structure of the magnetic field in the vicinity of the hot Jupiter
HD~209458b, for which we carried out numerical modeling in this paper. The
parameters of the magnetic field (the strength of the field and the orientation
of the magnetic axis) of the planet correspond to those set in the calculations
(see \S\ref{sec-res}). In  Fig.~\ref{fg2a} we show the distribution of
magnetic field lines for the case $B_0=10^{-3}$~G, which corresponds to the weak
wind field. The star is to the left and the planet is to the right. The star is
indicated by a color ring. The inner radius of the ring corresponds to the stellar
surface, while the outer radius --- to the surface of the corona. The radius of
the corona is about three times larger than the radius of the star. Thick solid
line labels the boundary of the Roche lobe. Magnetic field lines are shown by
solid lines with arrows. It is easy to see that the magnetic field can be
clearly divided into four magnetic zones, which are labeled by the corresponding
numbers. Zone 1 is characterized by the open stellar magnetic lines; 
magnetic lines originate at the surface of the star and extend to infinity. Zone
2 is defined by the similar open magnetic field lines of the planet. In zone
3, magnetic lines are common to the star and the planet, they originate at the
surface of the star and terminate at the surface of the planet. Finally, zone 4
consists of the closed lines of the planet field. At neutral points the
direction of the magnetic field is undefined. In the equatorial plane, these
points are labeled as $N_1$ and $N_2$. In the space, the set of these points
forms a neutral line similar to a circle, with the shape determined by the
orientation of the magnetic axis of the planet.
 
In Fig.~\ref{fg2b} we show the distribution of the magnetic field lines for the case
$B_0=1$~G, which corresponds to a strong wind field. Like in the previous case,
one can also distinguish all four magnetic zones and to define position of the
neutral points (see the right panel of the Figure). It should be noted, that
such a situation is not at all exotic, since such a field strength must be
typical for the stars of this type (spectral type of HD~209458 is G0\,V). For
instance, it is well known that the average magnetic field at the surface of the
Sun (including spots) is, approximately, 1~G.

\section{Magnetospheres of the hot Jupiters}

Under assumption of a constant radial velocity of the stellar wind $v_r = v_w$, it is
possible to derive a simple expression for the Alf\'{v}en Mach number:
\begin{equation}\label{eq-ms1}
 \lambda^2 = \frac{4\pi\rho_w v_w^2}{B_w^2} \left( \frac{r}{A} \right)^4 = 
 \lambda_w^2 \left( \frac{r}{A} \right)^4,
\end{equation}
where 
\begin{equation}\label{eq-ms2}
 \lambda_w = \frac{\sqrt{4\pi\rho_w} v_w}{B_w}
\end{equation}
defines the value of $\lambda$ at the orbit of the planet. At this,  
Alf\'{v}en point is defined by the expression
\begin{equation}\label{eq-ms3}
 a = \frac{A}{\sqrt{\lambda_w}}.
\end{equation}

In the solar wind,  Alf\'{v}en radius is \cite{Baranov1977}
\begin{equation}
\label{eq-ms4}
 a = 0.1~\text{AU} = 22 R_{\odot}.
\end{equation}
Since the semi-major axis of the orbit of the innermost planet, Mercury, is
$0.38~\text{AU} = 82 R_{\odot}$, this means that all planets of the Solar
System are located in the super-Alf\'{v}en zone of the solar wind. In the solar
wind, the sonic point, where the wind velocity becomes equal to the sound velocity,
is even closer to the Sun, at the distance of approximately $0.05\text{AU} = 11
R_{\odot}$. Then the magnetospheres (if any) of all planets
in the Solar System have a similar structure, resembling that of the Earth
magnetosphere. They are characterized by the following set of the basic
elements: bow shock, transition region, magnetopause, radiation belts,
magnetospheric tail.

In the case of hot Jupiters, because of their proximity to the parent star, the
structure of the magnetosphere may be completely different. Let consider as an
example two typical hot Jupiters HD~209458b and WASP~12b. For the first planet
one has: $A = 10.2 R_{\odot}$, $B_w = 0.0125$ G, $\lambda_w = 0.37$, $a = 16.8 R_{\odot}$.
At this, at the orbit of the planet  the ratio $ B_{\varphi}/B_r = 0.12$. For another
planet $A = 4.9 R_{\odot}$, $B_w = 0.1$ G, $\lambda_w = 0.045$, $a = 23.2 
R_{\odot}$  and the ratio of the azimuthal field and the radial one at the orbit 
of the planet is $B_{\varphi}/B_r = 0.01$. Thus,  these hot Jupiters
are located in the sub-Alf\'{v}en zone of the stellar wind. Accounting for the 
orbital motion may partially change the situation. In fact, the full wind velocity
relative to the planet in this case will be equal to 
$v = \sqrt{v_r^2 + v_{\varphi}^2} $, where $v_{\varphi} = \Omega A $, 
$\Omega = \sqrt{G M/A^3}$ is the orbital angular velocity of the planet, 
$G$ is gravity constant, $M = M_p + M_s$ --- the total mass of the system, 
$M_p$ --- the mass of the planet, $M_s$ --- the mass of the star.  Substituting 
the values of the corresponding parameters at the  orbits, we find
the ratio $v/u_A = 0.65$ for the planet HD~209458b and $v/u_A = 0.11$ for
the planet WASP~12b. As it can be seen, in the first case, the value of the 
total wind velocity turns out to be quite close to the Alf\'{v}en velocity. 
Therefore, it is possible 
to consider that the planet HD~209458b is located in the border region between
sub-Alf\'{v}en and super-Alf\'{v}en zones of the wind, since even small
magnetic field fluctuations (within a factor of 1.5 to 2) will be sufficient 
to change the mode of the wind flow around the planet.  

Because for these hot Jupiters Alf\'{v}en Mach number $\lambda = v_r/u_A$ turns
out to be less than 1, the ratio $v_r/u_F$, where $u_F = \sqrt{c_s^2 + u_A^2}$
and $c_s$ --- the sound velocity, will also be less than 1, since it is obvious
that $u_F > u_A$ and, therefore, the ratio $v_r/u_F < v_r/u_A$. In other words,
in the neighborhood of a hot Jupiter, stellar wind velocity will be lower than the
fast magneto-sonic velocity. In usual gas dynamics this case corresponds to the
subsonic flow around a body, in which the bow shock does not form. Thus, we
arrive to the following conclusion: the flow of stellar wind around such a hot
Jupiter should be \textit{shock-less}. In the structure of the magnetosphere of
the hot Jupiter bow shock should be absent.

This conclusion is based on the analysis of the parameters of two typical hot
Jupiters, HD~209458b and WASP~12b. However, apparently, it will remain valid for
many other exoplanets of this type. To verify this statement, we processed the
relevant data for a sample of 210 hot Jupiters taken from the database at
www.exoplanet.eu. The sampling was carried out by the masses of the planets
(mass of the planet $ M_p > 0.5 M_\text{jup}$, where $M_\text{jup}$ is the mass
of Jupiter), the orbital periods ($P_\text{orb} < 10$~day) and the semi-major
orbits ($A < 10 R_ {\odot}$). In addition, only those planets were kept for
which all necessary data are known.

\begin{figure}
\centering
\includegraphics[width = 0.49\textwidth]{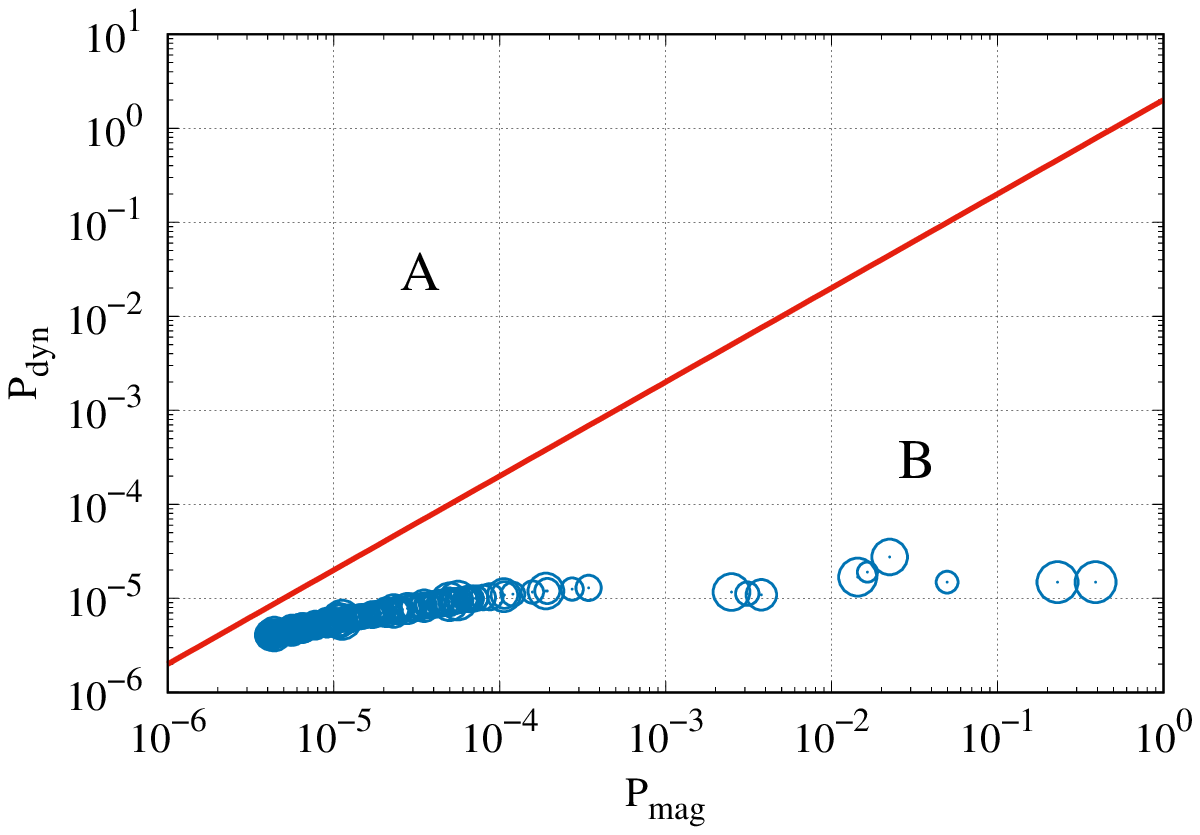}
\includegraphics[width = 0.49\textwidth]{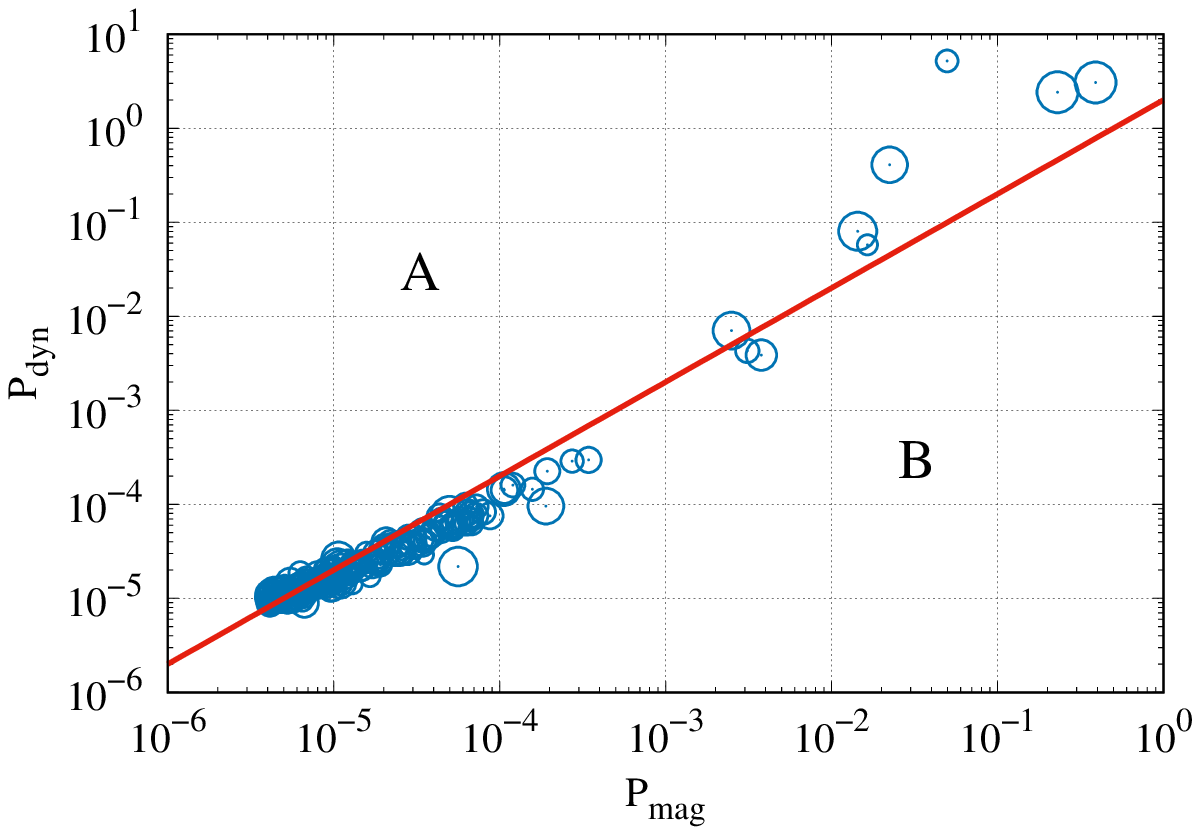}
\caption{Distribution of hot Jupiters in the two-dimensional diagram
$P_\text{mag}$ - $P_\text{dyn}$ (see the text for details). For the data in the
left panel, Alf\'{v}en  Mach numbers were calculated taking into account only
wind velocity. In the right panel computed data take into account orbital
velocities of the planets. The parameters of the planets are taken from the
database at www.exoplanet.eu. The data for 210 hot Jupiters was used. To positions
of the planets correspond the centers of the circles. The size of the circles
corresponds to the masses of the planets (in the logarithmic scale). Solid line
shows position of the Alf\'{v}en point. The letters label super-Alf\'{v}en zone
(A) and sub-Alf\'{v}en zone (B). (A color version of the Figure is available in
the electronic version of the journal).}%
\label{fg3}
\end{figure}

As a model of the stellar wind in the immediate vicinity of the Sun at the 
distances $1 R_{\odot} < r < 10 R_{\odot}$ we used results of the calculations from  
\cite{Withbroe1988}. According to the obtained profiles of density 
$\rho(r)$ and radial velocity $v_r(r)$, for every hot Jupiter from the sample we 
calculated the dynamic pressure of the wind at the orbit of the planet
\begin{equation}\label{eq-ms5}
 P_\text{dyn} = \rho(A) \left[ v_r^2(A) + \frac{G (M_s + M_p)}{A} \right]
\end{equation}
and magnetic pressure
\begin{equation}\label{eq-ms6}
 P_\text{mag} = \frac{B_r^2(A)}{8\pi},
\end{equation}
where the value of the radial field was calculated by the formula $B_r(A) =
B_0(R_\odot/A)^2$ with the parameter $B_0 = 1$~G. The resulting distribution of
hot Jupiters in the two-dimensional diagram $P_\text{mag}$ - $P_\text{dyn}$ is
presented in Fig.~\ref{fg3}. The left panel of the Figure presents results of
computations in which in calculation of the Alf\'{v}en Mach number only radial
wind velocity was taken into account. The right panel presents distribution of
planets for the case when their orbital velocity was taken into account. To the
positions of the planets correspond the centers of the circles; the sizes of the
latter are determined by their mass $M_p$ in the logarithmic scale. The solid
line shows position of the Alf\'{v}en point, to which corresponds a simple ratio
$P_\text{dyn} = 2 P_\text{mag}$. 
%Численное моделирование 
%обтекания газовой оболочки горячего юпитера плазмой звездного ветра обычно нами 
%производится в расчетной области, занимающей несколько десятков радиусов 
%планеты. На рис. \ref{fg3} закрашенная область (серая зона) соответствует 
%случаям, когда альфвеновская точка может попадать в расчетную область.  

As it is seen from the resulting distribution, many hot Jupiters from this
sample reside in the sub-Alf\'{v}en zone of the stellar wind. An account for the
orbital velocity substantially shifts  the entire sequence   upward in the diagram,
into direction of the super-Alf\'{v}en wind zone. Note that most of the planets
in this diagram form a certain regular sequence (see lower left corner of the
diagram). These planets are located quite far from the stars, where the
dependencies of density and wind velocity on the radius are well described by
power laws. Planets close to the stars are scattered over diagram in a rather
chaotic manner. For these planets, the magnitude of the dynamic wind pressure is
determined mainly by their orbital velocity. Note that the orbital velocity of
the planet depends not on the radius of the orbit only, but also (albeit to a
rather small extent) on the mass of the planet.
%Рис. \ref{fg3} показывает, что в серую зону могут попадают как далекие от звезды 
%горячие юпитеры (большая скорость ветра), так и близкие (большая орбитальная 
%скорость).

It should be borne in mind that this distribution was obtained for the solar
wind in the model of a quiet Sun. At this, we assumed that the average value of
the magnetic field at the surface of the Sun is 1~G. Even for the Sun, during
its activity cycle, positions of the hot Jupiters in the  diagram 
in Fig.~\ref{fg3} may
change in any direction with respect to the Alf\'{v}en point. In reality, every
planet of our sample is flown around not by the solar wind, but by the stellar
wind of the parent star. The parameters of this wind can significantly differ
from the solar wind ones. This means that the flow of the stellar wind around the
atmosphere of the planet must be investigated separately in each particular
case, taking into account the individual characteristics of the planet and the
parent star. In particular, in our numerical model, we can vary the value of the
average field $B_0$ at the surface of the star (i.e., at $r = R_s$, and not at
the surface of the Sun at $r = R_\odot $). The strength of the average magnetic
field of the solar type stars can range from about 0.1~G to about 5~G 
\cite{Fabbian2017}. In addition, the radii of the stars can be both smaller and
larger than the solar one. For example, the radius of the star WASP~12 is
1.57$~R_{\odot}$. Therefore, if one would  take the corresponding value of the
average field $B_0 = 1~\text{G}$, magnetic induction in the vicinity of the
planet WASP~12b will be, approximately, by factor 2.5 larger than the magnetic
induction of the solar wind at the same distance from the Sun. Using the same
simple method it is possible to model numerically formation  of magnetospheres 
of hot Jupiters of all major types.

Let characterize the magnetosphere by three characteristic parameters:
dimensions of \emph{ionospheric envelope} $R_\text{env}$, magnetopause radius
$R_\text{mp}$, and the radius of the bow shock $R_\text{sw}$. As the ionospheric
envelope, we mean the upper layers of the atmosphere of a hot Jupiter, which
consist of almost completely ionized gas \cite{Cherenkov2018}. In our
terminology, closed ionospheric envelope corresponds to the case when the
atmosphere of a hot Jupiter is entirely located within its Roche lobe. Open
ionospheric enelope corresponds to the case when a hot Jupiter overflows its
Roche lobe. The overflow results in planetary outflows from the vicinities of
the Lagrange points $L_1$ and $L_2$. For the magnetopause and shock wave, one
can take the distances from the center of the planet to the corresponding
frontal collision point. Depending on the relationship between these parameters,
we can suggest the following simple classification of the possible types of the 
magnetospheres of hot Jupiters.

Type \textbf{A}. The parameter $\lambda_w > 1$, therefore, 
a bow shock settles in the front of the magnetosphere, $R_\text{sw} < \infty$.
Taking into account
relations between the remaining parameters we obtain two special cases.
\begin{figure}
\centering
\includegraphics[width = 0.45\textwidth]{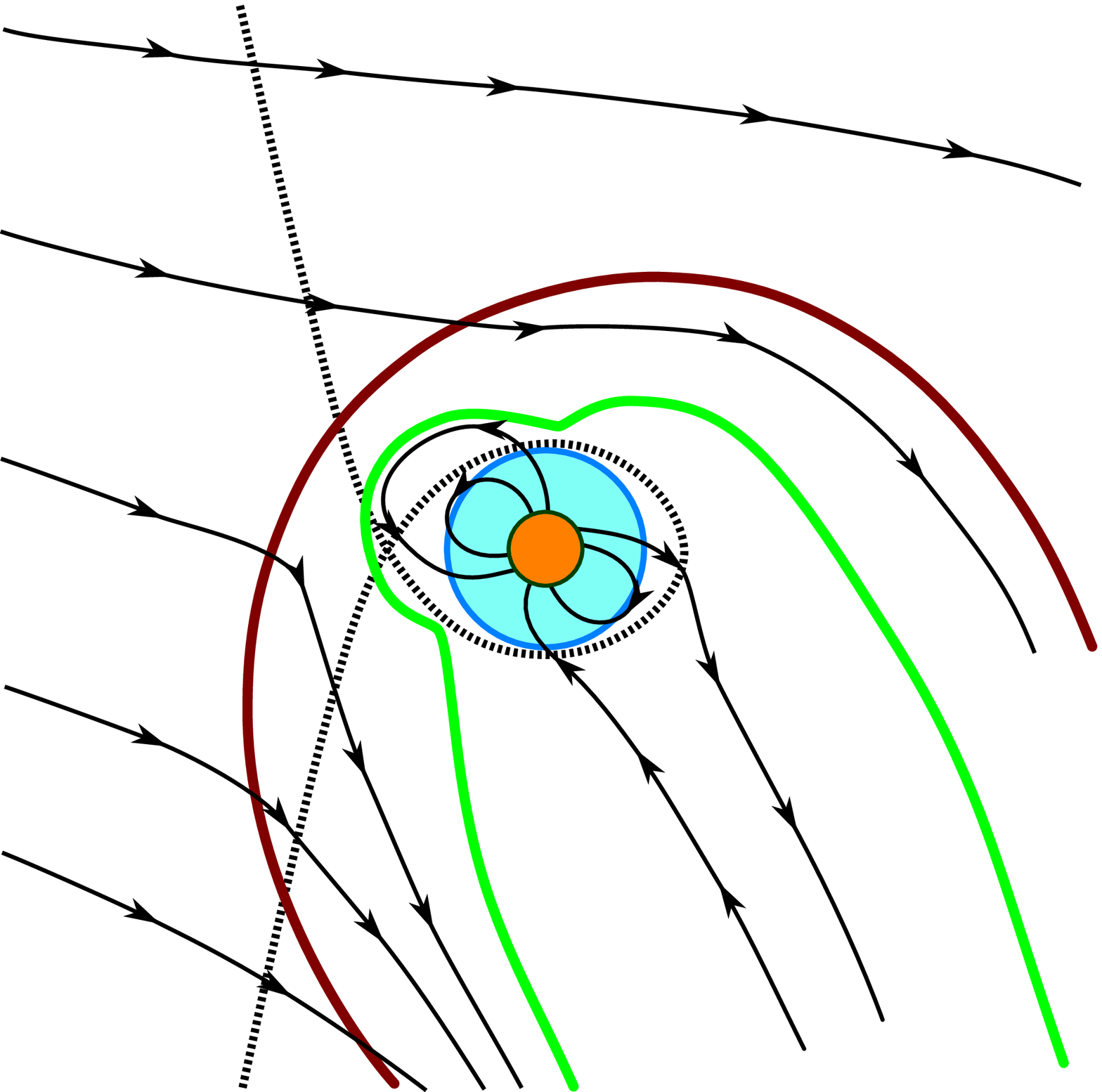}    %5
\includegraphics[width = 0.45\textwidth]{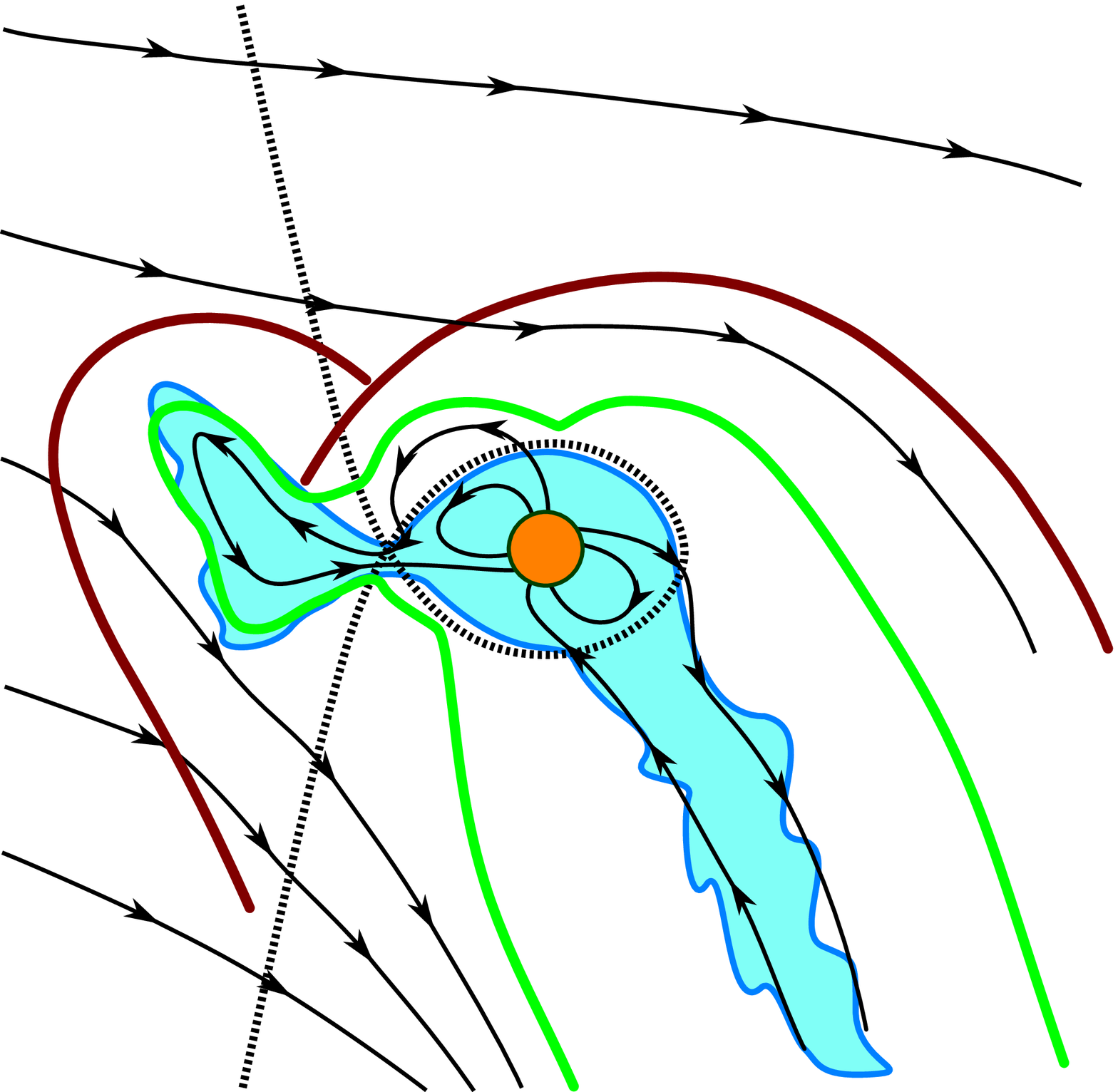}
\caption{Schematic representation of the structure of the 
 \textit{A1}-subtype  magnetosphere in the case of a closed (left) and open
(right) ionospheric envelope of a hot Jupiter. The lines with arrows correspond to the
magnetic field lines. Dotted line shows the border of the Roche lobe. The hatched
area corresponds to the gas envelope of the planet. Positions of the shock wave
(outer solid line) and magnetopause (inner solid line) are shown. (A color
version of the Figure is available in the electronic version of the journal).} %
\label{fg-a1}
\end{figure}

Subtype \textit {A1} (\textit{intrinsic magnetosphere
with bow shock}): $R_\text{env} < R_\text{mp}$. 
In this case, magnetic field of the planet is rather strong,
therefore, the magnetopause is located outside the ionospheric envelope. Corresponding
scheme of the structure of such a magnetosphere for the cases of closed and open ionospheric
envelopes is shown in Fig.~\ref{fg-a1}. In the solar system, similar situation for a
closed ionospheric envelope corresponds, for instance, to the magnetospheres of the Earth and
Jupiter.
\begin{figure}   %6
\centering
\includegraphics[width = 0.45\textwidth]{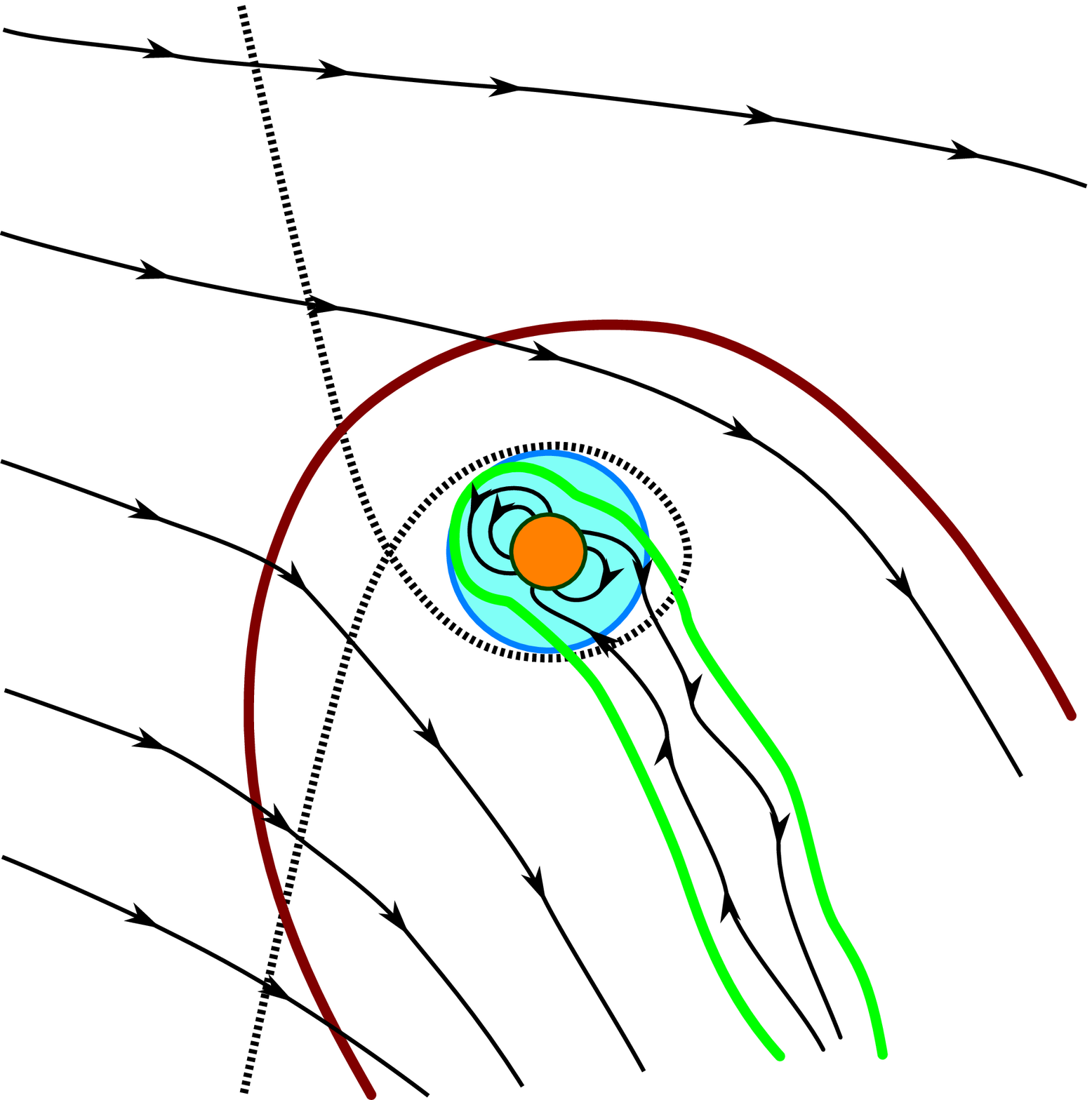}
\includegraphics[width = 0.45\textwidth]{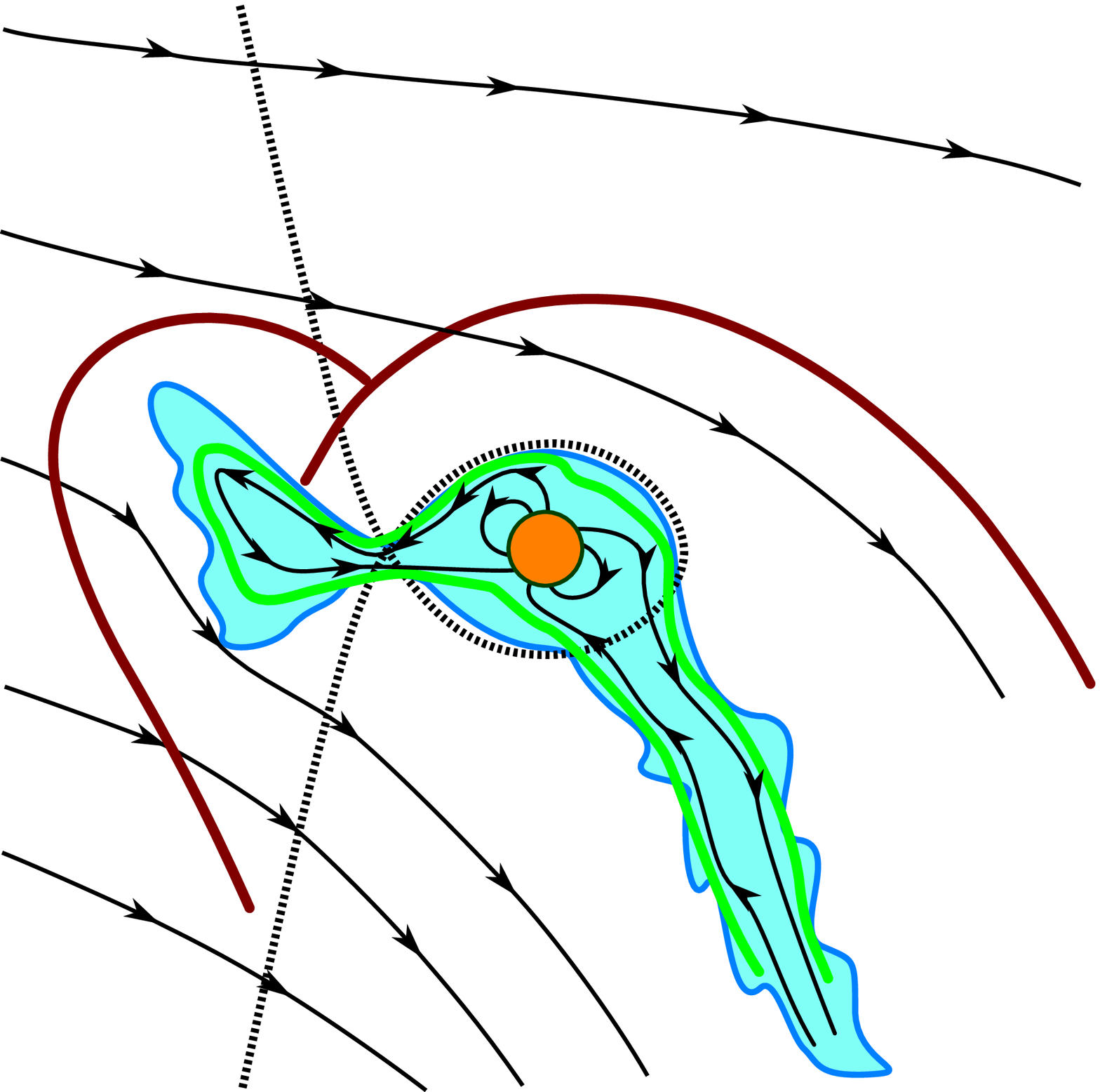}
\caption{Schematic representation of the structure of \textit{A2}-subtype
magnetosphere of in the the case of a closed (left) and open (right) ionospheric
envelope of a hot Jupiter. Notation as in Fig.~\ref{fg-a1}. (A color version of
the Figure is available in the electronic version of the journal.)}%
\label{fg-a2}
\end{figure}

Subtype \textit{A2} (\textit{induced magnetosphere
with bow shock}): $R_\text{env} > R_\text{mp}$. 
In this case, magnetic field of the planet is weak and, therefore,
the magnetopause is located inside the ionospheric envelope. Schematically, the
structure of such a magnetosphere for the cases of closed and open ionospheric
envelopes is shown in Fig.~\ref{fg-a2}. In the Solar System, this situation for
the case of a closed ionospheric envelope corresponds to the Venus magnetosphere
(and, to some extent, to the Mars one). 

Induced magnetosphere \cite{Russell1993} is formed by the currents that are
excited in the upper layers of the ionosphere. Excitation mechanism of these
currents is associated with the phenomenon of unipolar induction \cite{Landau8},
arising when the conductor moves perpendicularly to the magnetic field. The
currents induced in the ionosphere partially shield the magnetic field of the
wind. As a result, magnetic lines of the arising field enshroud the ionosphere
of the planet, forming a peculiar magnetic barrier (\emph{ionopause}). Bow shock
sets directly in the front of this barrier. On the night side, a magnetospheric
tail is formed, which can be partially filled by plasma from the ionosphere.
Unlike the proper magnetosphere, the orientation of the magnetic field in the
induced magnetosphere is completely determined by the wind field. As a result,
entire structure of the magnetosphere will track the direction to the star as
the planet moves along its orbit.

Type \textbf{B}. The parameter $\lambda_w < 1$ and the bow shock does not form.
Therefore, we can formally assume that $R_\text{sw} = \infty$. Again,
two particular cases can be distinguished.

\begin{figure}     %7
\centering
\includegraphics[width = 0.45\textwidth]{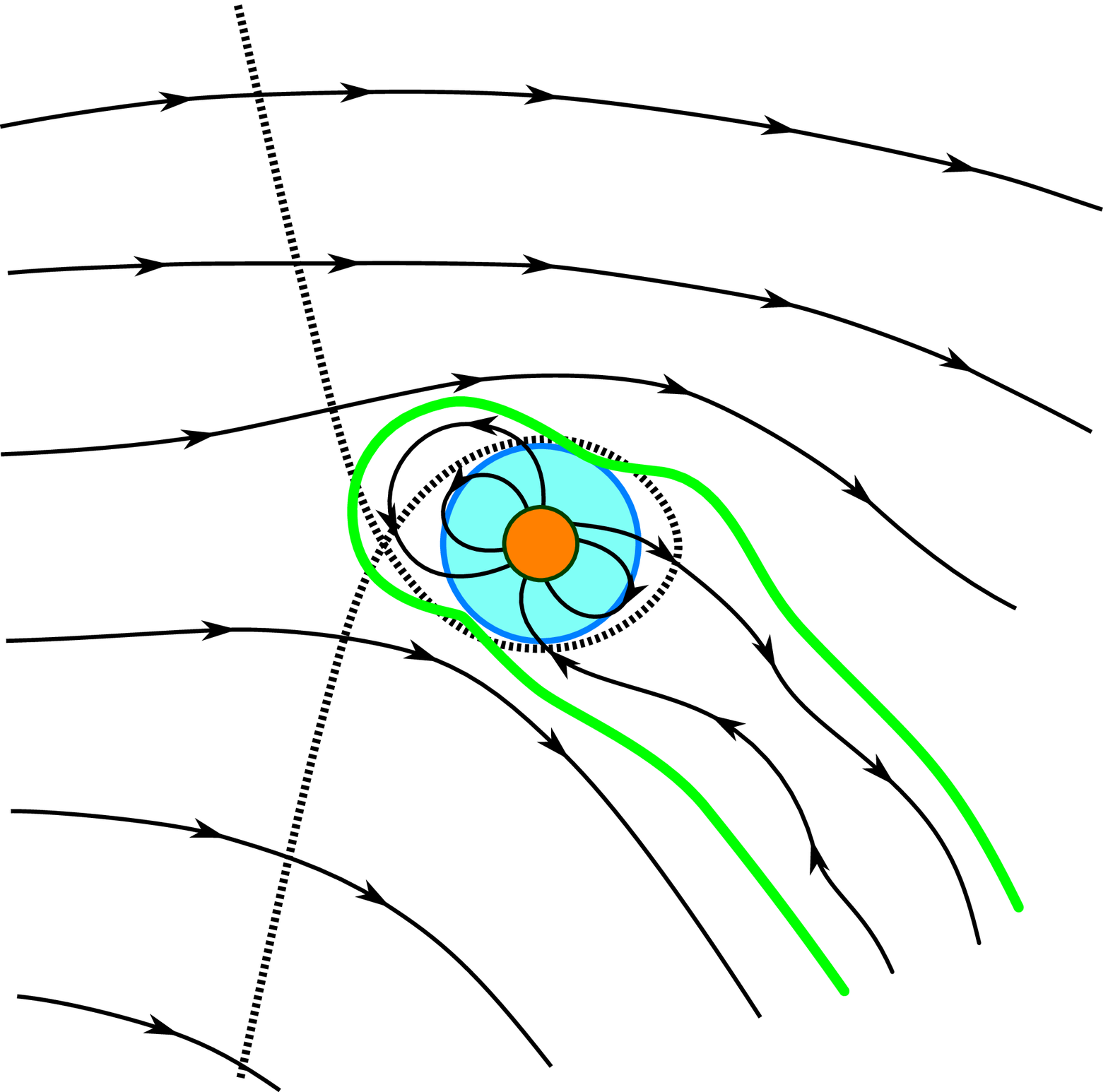}
\includegraphics[width = 0.45\textwidth]{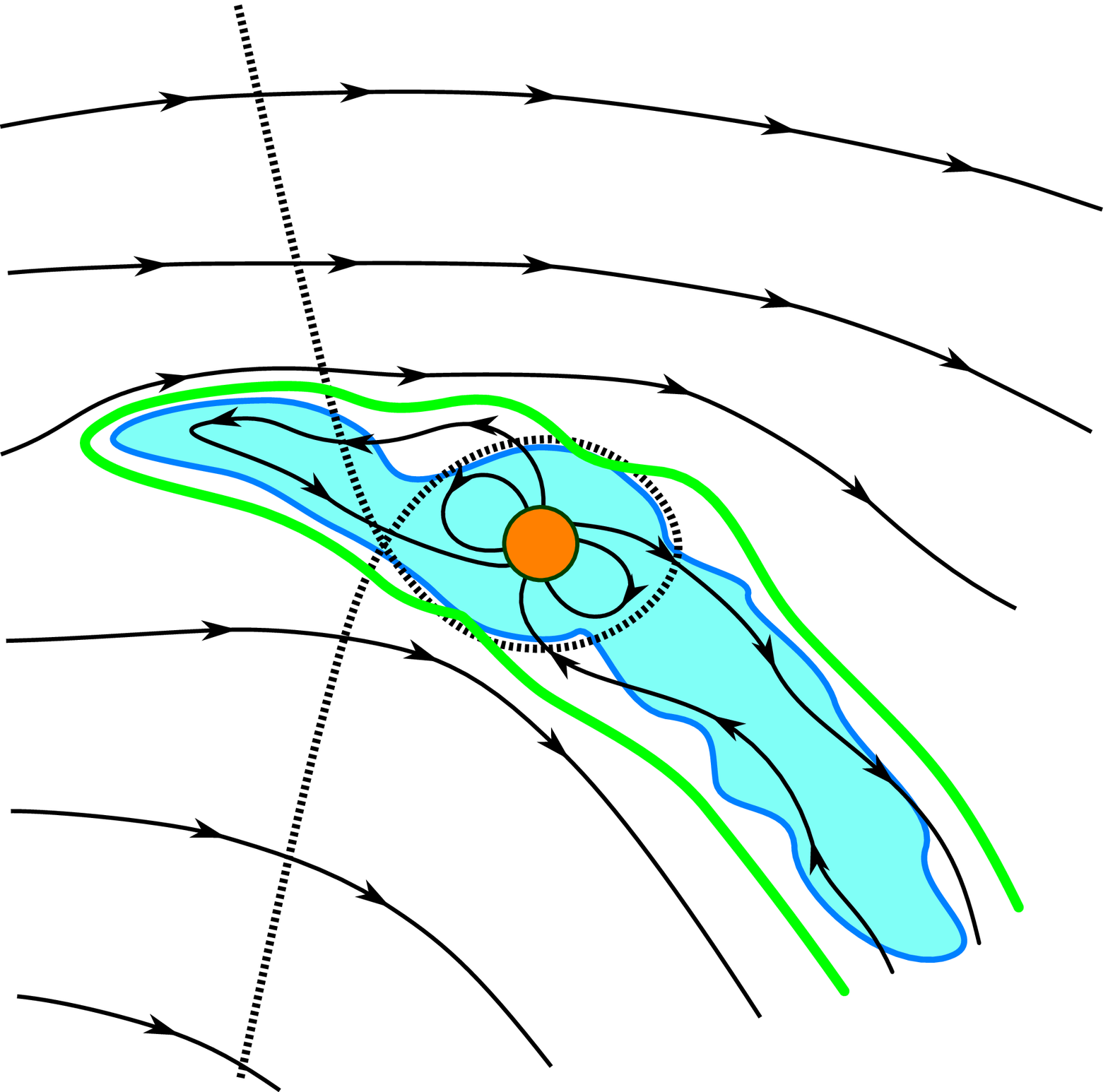}
\caption{Schematic representation of the structure of \textit{B1}-subtype
magnetosphere 
in the the case of a closed (left) and open (right) ionospheric
envelope of a hot Jupiter. Notation as in Fig.~\ref{fg-a1}. (A color version of
the Figure is available in the electronic version of the journal.) }%
\label{fg-b1}
\end{figure}

Subtype \textit{B1} (\textit{shock-less intrinsic
magnetosphere}): 
$R_\text{env} < R_\text {mp}$. This situation arises in the case of a sufficiently
strong own magnetic field of the planet. As a result, the boundary of the magnetopause
will be located outside the ionospheric envelope. The structure of the magnetosphere
of this type for the cases of closed and open ionospheric envelopes are shown in 
Fig.~\ref{fg-b1}. It should be noted that, apparently, this case is rather
exotic, because the own magnetic field  of a hot Jupiter must be relatively weak.
\begin{figure} %8
\centering
\includegraphics[width = 0.45\textwidth]{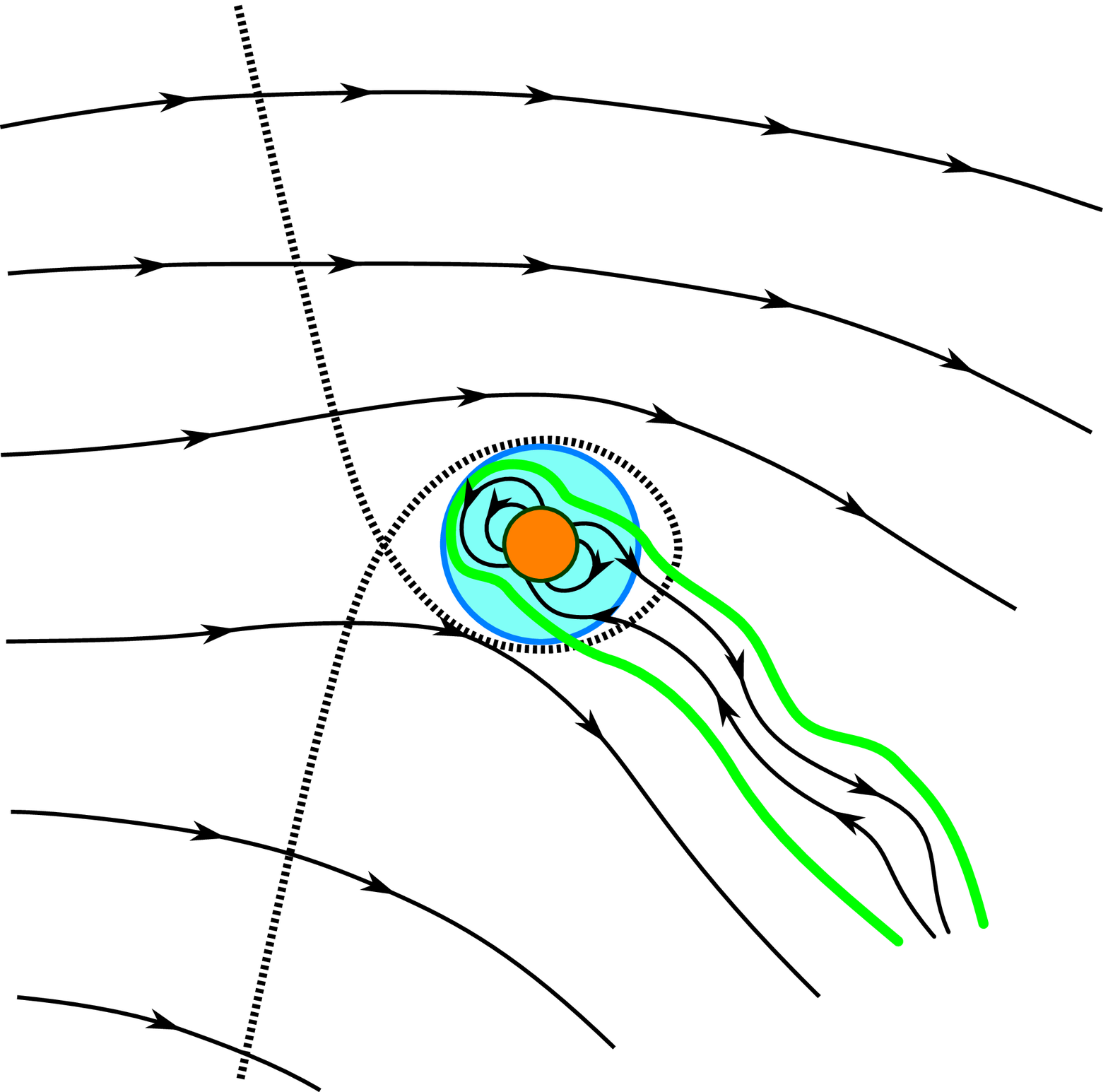}
\includegraphics[width = 0.45\textwidth]{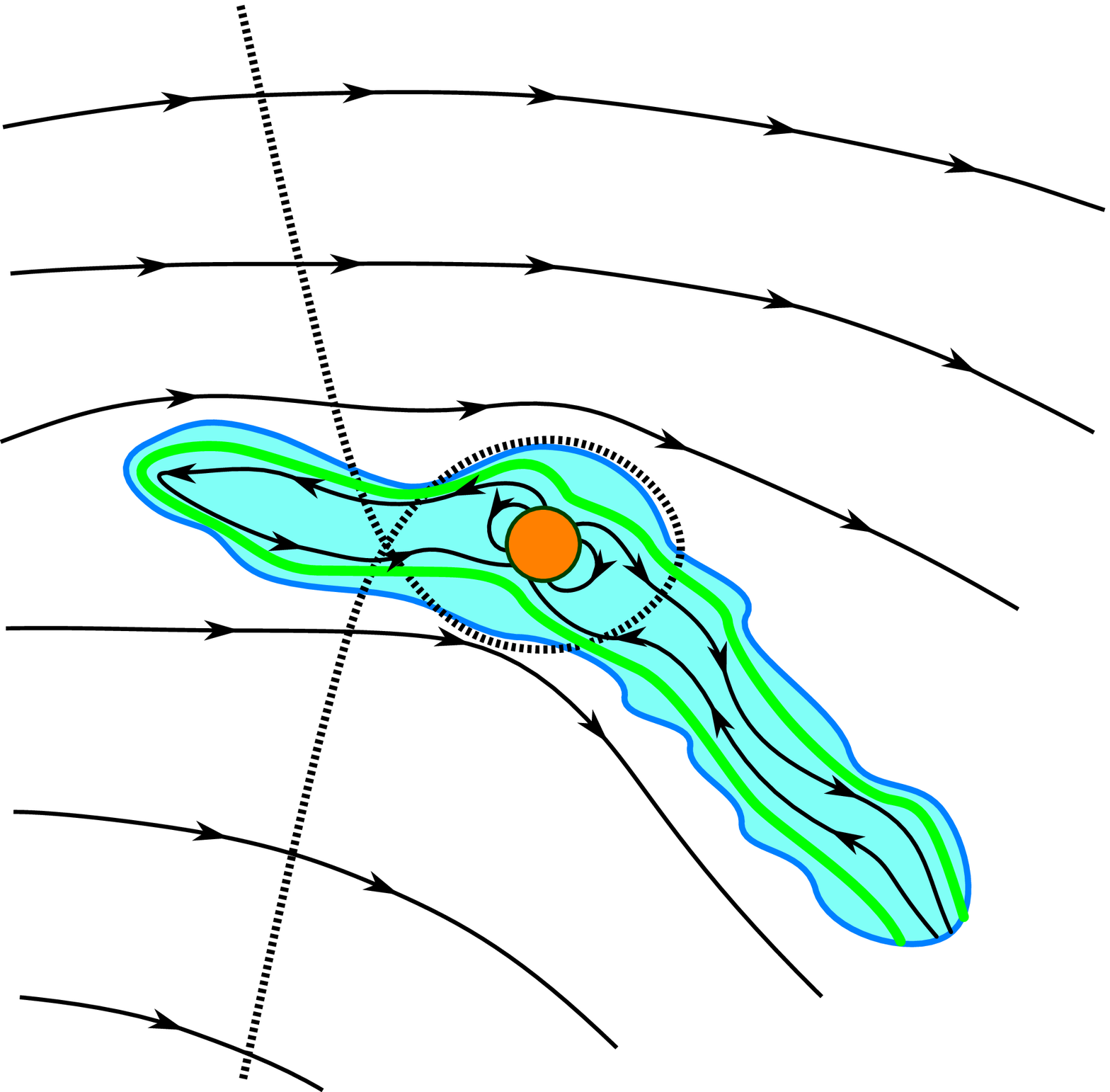}
\caption{Schematic representation of the structure of \textit{B2}-subtype
magnetosphere  
in the case of a closed (left) and open (right) ionospheric
envelope of a hot Jupiter. Notation as in Fig.~\ref{fg-a1}. (A color version of
the Figure is available in the electronic version of the journal.)}%
\label{fg-b2}
\end{figure}

Subtype \textit{B2} (\textit{Shock-less induced magnetosphere}): $ R_\text{env} >
R_\text{mp}$. It is possible that for the hot Jupiters this is the most common
situation. In this case, the magnetopause is formally located inside the
ionospheric envelope. Therefore, the outflows from the envelope interact
directly with the magnetic field of the stellar wind. The structure of the
magnetosphere of this type for the cases of closed and open ionospheric
envelopes is shown in Fig.~\ref{fg-b2}. Taking into account possible types of
gas envelopes of hot Jupiters \cite{Bisikalo2013}, it is possible to distinguish
here additional subtypes, corresponding, for instance, to closed, quasi-closed,
and open envelopes..

Type \textbf{C}. Parameter $\lambda_w \approx 1$. This is an intermediate type
of magnetospheres corresponding to the ``gray'' zone. In particular, in this
case the planet itself can be located in the sub- or super-Alf\'{v}en wind zone
at the time, when outflowing ionospheric envelope, due to its rather large
extent, may intersect the Alf\'{v}en point and partially flow into the opposite
wind zone. Such an unusual situation may appear fairly common for hot Jupiters,
since their orbits are usually located close to the Alf\'{v}en point (see the
distribution of hot Jupiters in the diagram in Fig.~\ref{fg3}). This case must be
investigated separately.

\section{Description of the model}

\subsection{Basic equations}

To describe the structure of the flow in the vicinity of hot Jupiter, we will
use the system of equations of ideal one-fluid magnetic hydrodynamics with
background field \cite{mcb-book, zbbUFN, Arakcheev2017}. Under this approach,
the full magnetic field $\vec{B}$ is represented as a superposition of the
background magnetic field $\vec{H}$ and the magnetic field $\vec{b}$ induced by
the currents in the plasma itself, $\vec{B} = \vec{H} + \vec{b}$. Since the
background field in the problem in question is generated by the sources outside
the computational domain, in the the computational domain it must satisfy the
potential condition, $\nabla \times \vec{H} = 0$. Just this external field
property is used for its partial exclusion from the equations of magnetic
hydrodynamics \cite{Tanaka1994, Powell1999}. In addition, in our model we assume
that the background magnetic field is steady, $\spdiff{\vec{H}}{t} = 0$. This
corresponds to the case, when the spin of a hot Jupiter's is synchronized with its
orbital motion.

Bearing in mind condition 
$\nabla \times \vec{H} = 0$, equations of ideal magnetohydrodynamics may be 
written as 
\begin{equation}\label{eq-mhd1}
 \pdiff{\rho}{t} + \nabla \cdot \left( \rho \vec{v} \right) = 0,
\end{equation}
\begin{equation}\label{eq-mhd2}
 \rho \left[ 
 \pdiff{\vec{v}}{t} + \vec{v} \cdot \nabla \vec{v} 
 \right] = 
 -\nabla P - \vec{b} \times \nabla \times \vec{b} - 
 \vec{H} \times \nabla \times \vec{b} - \rho \vec{f},
\end{equation}
\begin{equation}\label{eq-mhd3}
 \pdiff{\vec{b}}{t} = 
 \nabla \times \left( 
 \vec{v} \times \vec{b} + 
 \vec{v} \times \vec{H} 
 \right),
\end{equation}
\begin{equation}\label{eq-mhd4}
 \rho \left[ 
 \pdiff{\varepsilon}{t} + \vec{v} \cdot \nabla \varepsilon
 \right] + P\, \nabla \cdot \vec{v} = 0.
\end{equation}
Here$ \rho$ is density, $\vec{v}$ --- velocity, $P$ --- pressure, 
$\varepsilon$ --- specific internal energy. For convenience of numerical modeling, in these
equations the system of units without factor  
$4\pi$ is used. It is assumed that the matter may be considered 
as an ideal gas with an equation of state  
\begin{equation}\label{eq-mhd5}
 P = (\gamma - 1) \rho \varepsilon,
\end{equation}
where $\gamma = 5/3$ --- adiabatic exponent. 

Computations were carried out in a rotating reference frame, in which the positions 
of the centers of star and planet did not change. In this case, the angular 
velocity vector of rotation of the reference frame $\vec{\Omega}$ coincides with
the orbital angular velocity of the ``star - planet'' binary system. In such a
rotating reference frame, the specific external force is determined by the expression
\begin{equation}\label{eq-mhd6}
 \vec{f} = -\nabla \Phi - 2 \left( \vec{\Omega} \times \vec{v} \right).
\end{equation}
Here the first term in the right-hand side describes the force due to the Roche potential
gradient 
\begin{equation}\label{eq-mhd7}
 \Phi = -\frac{G M_s}{|\vec{r} - \vec{r}_s|} - 
 \frac{G M_p}{|\vec{r} - \vec{r}_p|} - 
 \frac{1}{2} \left[ \vec{\Omega} \times (\vec{r} - \vec{r}_c) \right]^2,
\end{equation}
where $M_s$ is the mass of the star, $M_p$ --- mass of the planet, $\vec{r}_s$
--- radius-vector of the stellar center, $\vec{r}_p$ --- radius-vector of the
center of the planet, $\vec{r}_c$ --- radius-vector of the center of mass of the
system. The second term describes the Coriolis force. 

The background magnetic field was set as $\vec{H} = \vec{H}_p + \vec{H}_s$. 
The first term, $\vec{H}_p$ describes the proper magnetic field of the planet. 
It was assumed in our model that hot Jupiters have a dipole magnetic field,
\begin{equation}\label{eq-mhd8}
 \vec{H}_p = \frac{\mu}{|\vec{r} - \vec{r}_p|^3} \left[
 3 (\vec{d} \cdot \vec{n}_p) \vec{n}_p - \vec{d}
 \right],
\end{equation}
where $\mu$ is magnetic momentum, 
$\vec{n}_p = (\vec{r} - \vec{r}_p) / |\vec{r} - \vec{r}_p|$, $\vec{d}$ --- unit vector, 
directed along magnetic axis, 
the vector of magnetic momentum  $\vec{\mu} = \mu \vec{d}$. 
The second term, $\vec{H}_s$, describes the radial magnetic field of the stellar wind:
\begin{equation}\label{eq-mhd9}
 \vec{H}_s = \frac{B_0 R_s^2}{|\vec{r} - \vec{r}_s|^2} \vec{n}_s,
\end{equation}
where $R_s$ is stellar radius, while the vector
$\vec{n}_s = (\vec{r} - \vec{r}_s) / |\vec{r} - \vec{r}_s|$. 
It is not difficult to ascertain that such a background magnetic field satisfies the 
potential condition $\nabla \times \vec{H} = 0$. Thus, in our model at the initial
instant of time the proper magnetic field of the plasma $\vec{b}$ will be defined 
by the azimuthal component of the magnetic field \eqref{eq-wmf18} only. 

\subsection{Numerical method}

For the numerical solution of the equations of magnetic hydrodynamics, presented
in the previous section, we use a combination of Roe \cite{Roe1980} and Lax -
Friedrihs \cite{Lax1954, Friedrihs1954} difference schemes. Solution algorithm
has several successive stages resulting from the application of splitting over
physical processes. Suppose, we know the distribution of all quantities over
computational grid at the time $t^n$. Then, to get the values at the next time 
instant $t^{n + 1} = t^n + \Delta t$, let decompose the complete
system of equations \eqref{eq-mhd1} - \eqref{eq-mhd4} into two subsystems.

The first subsystem corresponds to the ideal magnetohydrodynamics with proper 
magnetic field of the plasma  $\vec{b}$ without account for the background 
magnetic field $\vec{H}$:
\begin{equation}\label{eq-nm1}
 \pdiff{\rho}{t} + \nabla \cdot \left( \rho \vec{v} \right) = 0,
\end{equation}
\begin{equation}\label{eq-nm2}
 \rho \left[ 
 \pdiff{\vec{v}}{t} + \vec{v} \cdot \nabla \vec{v} 
 \right] = 
 -\nabla P - \vec{b} \times \nabla \times \vec{b} - \rho \vec{f},
\end{equation}
\begin{equation}\label{eq-nm3}
 \pdiff{\vec{b}}{t} = \nabla \times \left( \vec{v} \times \vec{b} \right),
\end{equation}
\begin{equation}\label{eq-nm4}
 \rho \left[ 
 \pdiff{\varepsilon}{t} + \vec{v} \cdot \nabla \varepsilon
 \right] + P\, \nabla \cdot \vec{v} = 0.
\end{equation}

In our numerical model, to solve this system,
the Roe scheme \cite{Cargo1997, Kulikovsky2001} 
with Osher's incremental correction \cite{Osher1985} (see also the 
monograph \cite{mcb-book}) for
magnetohydrodynamical equations
was used.
The magnetohydrodynamical version of the Roe scheme was introduced in the code in such a way
that in the absence of a magnetic field ($\vec{b} = 0$) this scheme  exactly transforms
into Roe -- Einfeldt -- Osher scheme, which we used in  purely
gas-dynamical calculations \cite{Bisikalo2013}.

The second subsystem accounts for the influence of the background field:
\begin{equation}\label{eq-nm5}
 \rho \pdiff{\vec{v}}{t} = -\vec{H} \times \nabla \times \vec{b},
\end{equation}
\begin{equation}\label{eq-nm6}
 \pdiff{\vec{b}}{t} = 
 \nabla \times \left( \vec{v} \times \vec{H} \right).
\end{equation}
The first equation in this subsystem describes the influence of the electromagnetic
force, caused by the background field, while the second equation describes the
generation of magnetic field. It is assumed that at this stage the density
$\rho$ and specific internal energy $\varepsilon$ do not change. To solve the
second subsystem, Lax-Friedrihs scheme with TVD  (total variation
diminishing) boost was applied \cite{mcb-book}).

To clear the divergence of the magnetic field $\vec{b} $, we used the method of
generalized Lagrange multiplier \cite{Dedner2002}. The choice of this method
is due to the fact that the flow in the vicinity of the hot Jupiter is essentially
unsteady, especially in the cocurrent stream, forming the magnetospheric tail.

\section{Results of the modeling}
\label{sec-res}

As an example demonstrating the ideas presented in the paper, we computed 
the model of the flow structure in the vicinity of a hot Jupiter HD~209458b. It is the
first transient hot Jupiter, discovered in 1999 \cite{Charbonneau2000}. The main
parameters of the model corresponded to the values used in our
previous studies (see, e.g., \cite {Bisikalo2013}). Parent star
has spectral type G0, the mass $M_s = 1.15 M_\odot$, the radius
$R_s = 1.2 R_\odot$. Proper stellar rotation is characterized by a period
$P_\text{rot} = 14.4$~ day, which corresponds to the angular velocity $\Omega_s = 
5.05 \cdot 10^{-6}~\text{s}^{-1}$ 
or the linear velocity at the equator
$v_\text{rot} = 4.2~\text{km}/\text{s} $. The mass of the planet is 
$M_p = 0.71~M_\text{jup}$, its photometric radius $R_p = 1.38~R_\text{jup}$, where
$M_\text{jup}$ and $R_\text{jup}$ are the mass and radius of the Jupiter, respectively.
Semi-major axis of the 
planet orbit $A = 10.2 R_\odot$, corresponding to the period of revolution $P_\text{orb} = 84.6$~hr.

At the initial instant of time we set a spherically-symmetric isothermal atmosphere with density
distribution defined by the following expression:
\begin{equation}\label{eq-res1}
 \rho = \rho_\text{atm} \text{exp} \left[
 -\frac{G M_p}{R_\text{gas} T_\text{atm}}
 \left(
 \frac{1}{R_p} - \frac{1}{|\vec{r} - \vec{r}_p|}
 \right)
 \right],
\end{equation}
where $\rho_\text{atm}$~--- the density at the photometric radius,
$T_\text{atm}$~--- the temperature of the atmosphere, 
$R_\text{gas}$~--- gas constant.

The radius of the atmosphere was determined from the condition of pressure equilibrium
with the matter of stellar wind. The following atmospheric parameters were used
in the calculations: temperature $T_\text{atm} = 7500~\text{K}$, 
the concentration of particles at photometric radius $n_\text{atm} = 10^{11}~\text{cm}^{-3}$.

As the parameters of the stellar wind, we used the corresponding parameters of
the solar wind at the distance $10.2R_\odot$ from the center of the Sun
\cite{Withbroe1988}: temperature $T_w = 7.3 \cdot 10^5~\text {K}$, 
velocity $v_w = 100~\text{km}/\text {s} $, concentration 
$n_w = 10^4~\text{cm}^{-3}$. Magnetic field of the wind was set
by the formulas presented in the description of the numerical model.

Kislyakova et al. \cite{Kislyakova2014} have  found from observational
data that the magnetic moment $\mu$ of the hot Jupiter HD~209458b can not exceed the
$0.1 \mu_\text{jup}$, where $\mu_\text{jup} = 1.53 \cdot 10^{30}~\text{G} \cdot 
\text{cm}^3$ is the magnetic moment of Jupiter. The estimate of the magnetic moment
of HD~209458b 
according to \cite{Stevenson1983} is, approximately, $ 0.08\mu_\text{jup}$.
In our calculations, we assumed that  the magnetic moment of the 
hot Jupiter HD~209458b is $\mu = 0.1\mu_\text{jup}$. Magnetic axis of the dipole
was tilted by the angle of $30^\circ$ relative to the axis of rotation of the planet 
in the direction 
opposite to the star.  At this, we considered that the proper rotation
of the planet is synchronized with the orbital rotation and the axis of its own
rotation is collinear with the axis of orbital rotation.

Computations were carried out in the Cartesian coordinate system, with the origin 
in the center of the planet. The $x$-axis connected the centers of the star and the
planet and was directed from the star. The $y$-axis was directed along the orbital
rotation of the planet, and the axis $z$~--- along its axis of proper rotation.
The computational domain had  dimensions 
$-30 \le x/R_p \le 30$, $-30 \le y/R_p \le 30$, $-15 \le z/R_p \le 15$,  the
number of cells was $N = 480 \times 480 \times 240$.   
To increase the spatial resolution in the atmosphere of the planet, we
used the grid, exponentially condensing to the center of the planet. Characteristic
size of the cell at the photometric radius of the planet was $0.02 R_p$,
while at the outer edge of the computational domain the cell size was equal,
approximately, to $0.4 R_p$. The boundary conditions were the same as in our
recent work \cite{Arakcheev2017}.

\begin{figure} %9
\centering
\includegraphics[width = 0.49\textwidth]{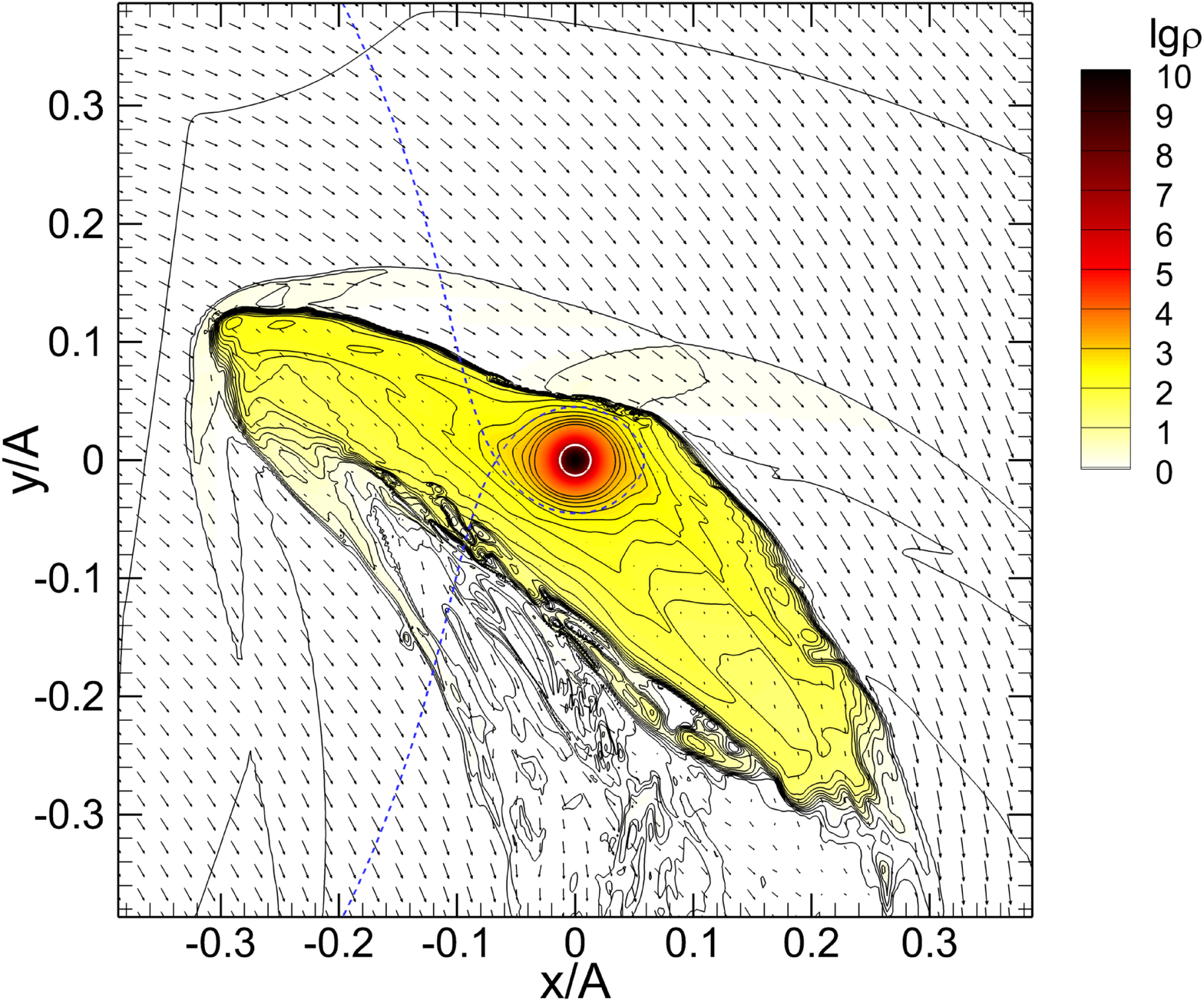}
\includegraphics[width = 0.49\textwidth]{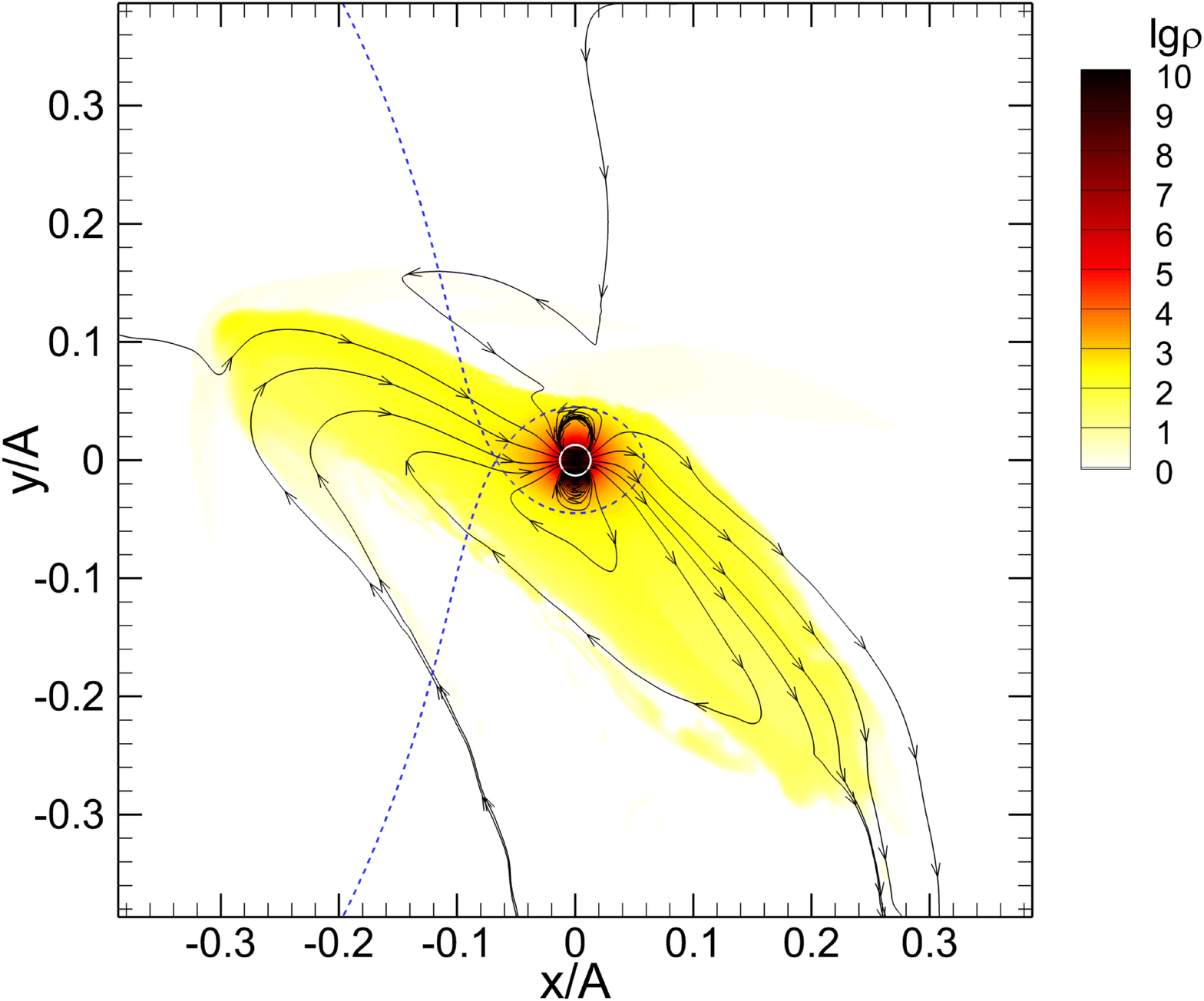}
\caption{ Distribution of density  (color 
scale and contours), velocity (the arrows, left panel), and magnetic field 
(the lines with arrows, right panel) in
the orbital plane of the hot Jupiter for the case of a weak magnetic field of the wind
(model 1). Solution is presented for the time $0.23 P_\text{orb}$ from the beginning
of computations. The dashed line shows the boundary of the Roche lobe. Light circle
corresponds to the photometric radius of the planet. (Color version of the Figure is
available in the electronic version of the journal.)}%
\label{fg-res1}
\end{figure}

\begin{figure}    %10
\centering
\includegraphics[width = 0.49\textwidth]{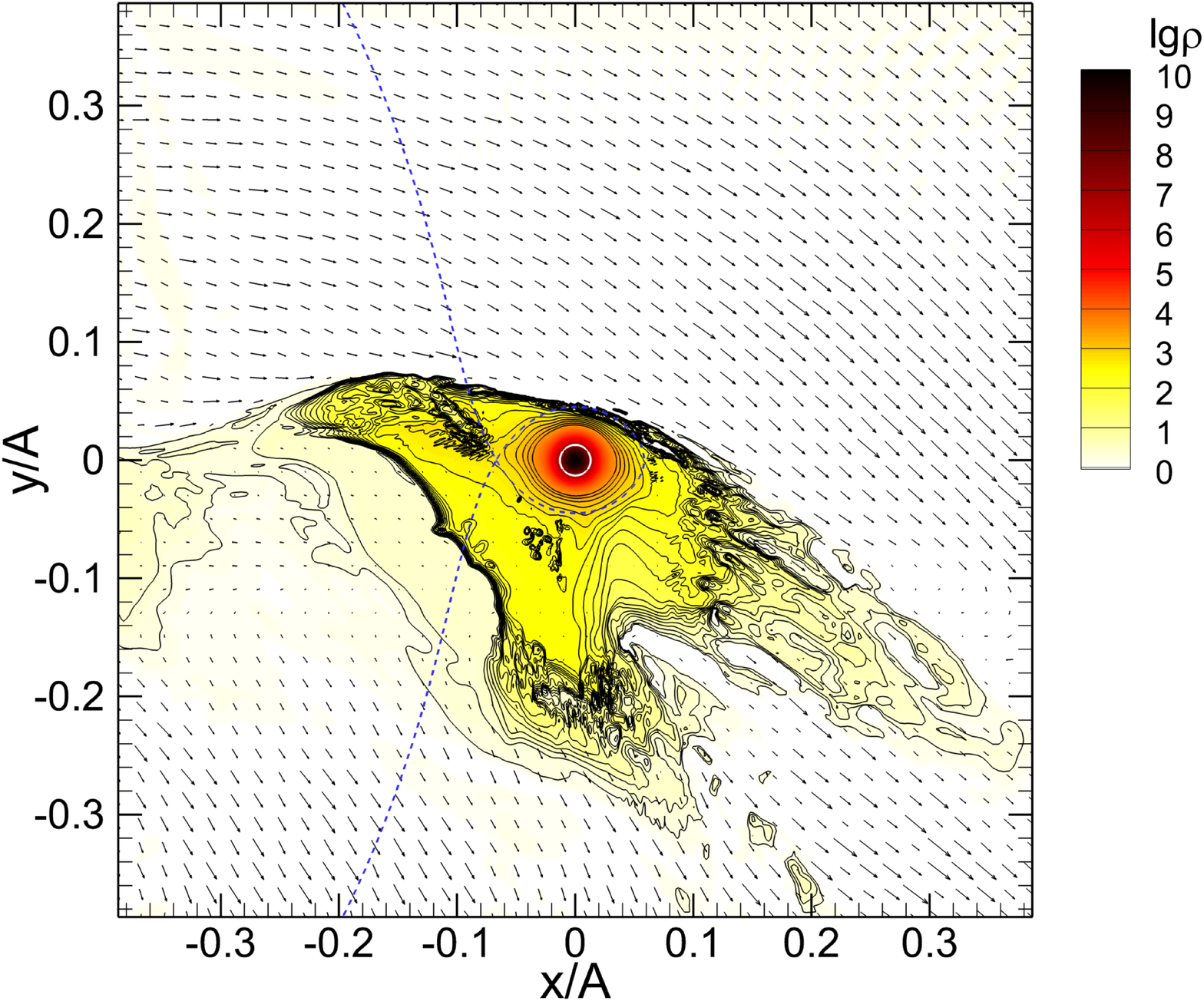}
\includegraphics[width = 0.49\textwidth]{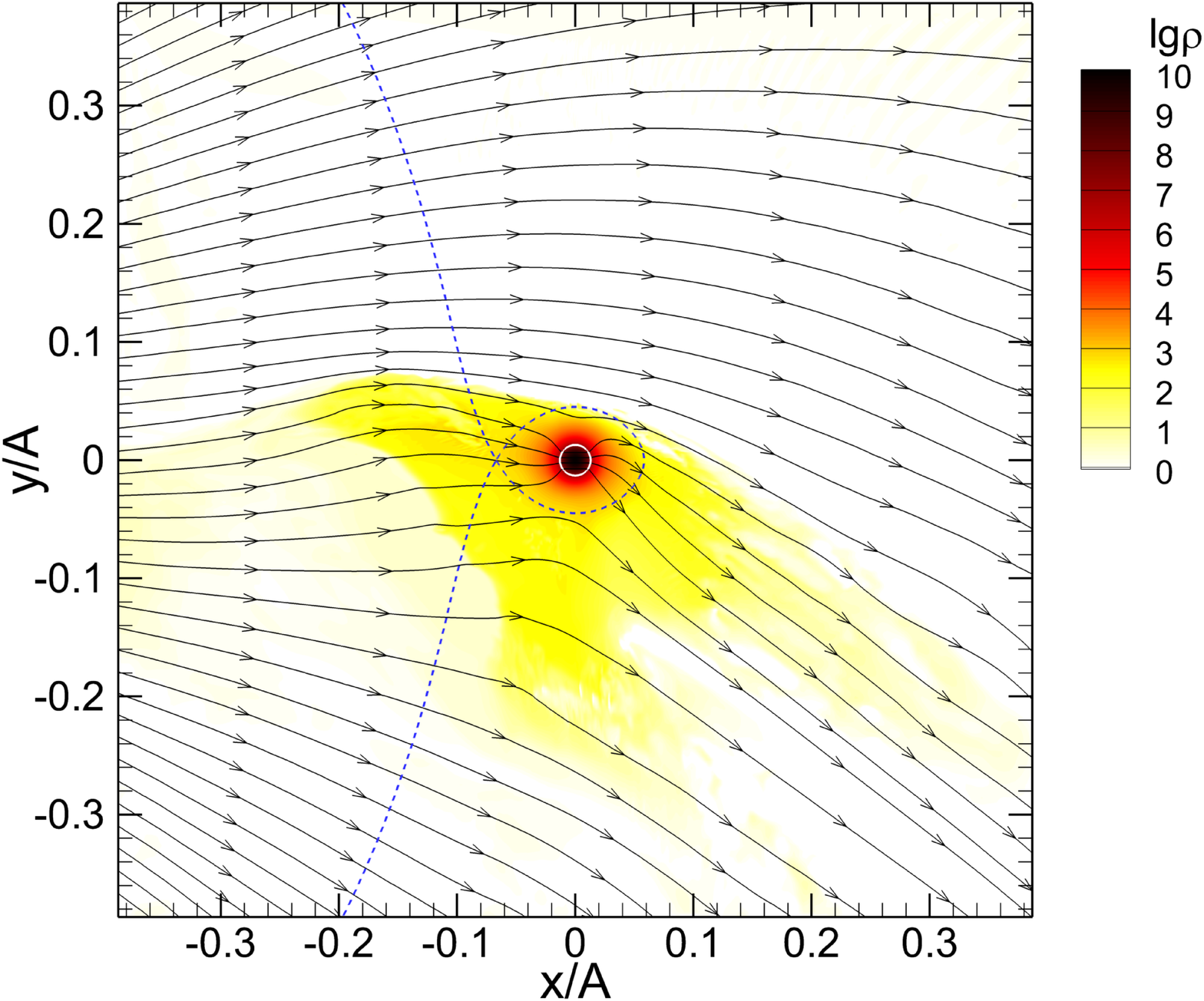}
\caption{Same as in Fig.~\ref{fg-res1}, but for the model 2.}%
\label{fg-res2}
\end{figure}

We carried out two calculations, which differed only in the value of the parameter
$B_0$, which determines the average magnetic field at the surface of the star. 
In the first run (model 1),  $B_0$ was set to $10^{-4}~\text{G}$. This
corresponds to a weak magnetic field of the stellar wind. In the second run
(model 2), $B_0 = 1~\text{G}$ was used (strong field). This  
corresponds to the average magnetic field of the quiet Sun. Results
of calculations are presented in Figs.~\ref{fg-res1} and \ref{fg-res2}. The Figures 
show density distribution (by gradations of color and contours), velocity (the 
arrows in the panels to the left), and magnetic field (the lines with the arrows in
the panels to  the right) in
the orbital plane of a hot Jupiter. The density is normalized to the 
wind density at the orbit of the planet $\rho_w$. Presented numerical solutions
correspond to the time $0.23 P_\text{orb}$ from the start of the computations. The border
of the Roche lobe is shown by a dotted line. The planet is located in the center of the 
computational domain and is depicted by a light circle, the radius of which corresponds to
the photometric radius.

In both models two powerful streams form from the neighborhood of the Lagrange points
$L_1$ and $L_2$. The first stream is formed on the day side, 
it is directed toward the star
and, therefore, it moves against the wind under the influence of its gravity. The 
second stream begins on the night side and forms a wide turbulent plume behind the planet.

In model 1, as a result of the interaction of the stellar wind with the envelope
of the planet, a detached shock wave forms, well visible in Fig.~\ref{fg-res1}. 
One can say that it consists of two separate shock waves, one of which
arises around the atmosphere of the planet, while the other~--- around the jet
from the inner Lagrange point $L_1$. In the right panel of Fig.~\ref{fg-res1} it is
seen that inside Roche lobe of the planet magnetic field remains close to the dipole.
However, in the outflows magnetic field lines are drawn by plasma flows. Magnetic field
of the stellar wind in this model is so weak that it plays no dynamic role. In fact, 
it manifests itself as a kind of a passive impurity present in the wind plasma. 
Such a magnetosphere obviously
corresponds to the \textit{A1} subtype in the case of an open ionospheric envelope
of the hot Jupiter, the structure of which is schematically shown in the right panel of 
Fig.~\ref{fg-a1}.

In the model 2, the interaction of the stellar wind with the envelope of the planet is
shock-less. In the left panel of Fig.~\ref{fg-res2} it is  seen that detached shock
does not form neither around the atmosphere of the planet nor around the jet from
the point $L_1$. Strong magnetic field of the wind impedes free motion of the matter
in the direction transverse to the lines of force. Therefore, the flow pattern in this
model differs significantly from that in the model 1, because in this case, apart of
stellar gravity, the centrifugal force, and Coriolis force, an 
essential role also belongs
to the electromagnetic force,  due to the magnetic field of the wind.  The tail by 
the planet is also oriented at a different angle, since the flows in it are aligned 
mainly along the magnetic field lines. Magnetic field of the wind is slightly
distorted by the streams from the planet (see the right panel in Fig.~\ref{fg-res2}),
but, generally, retains its original structure.
The magnetosphere in model 2 corresponds to the \textit{B2} subtype in the case of open
ionospheric envelope of the hot Jupiter; its structure is shown schematically
in the right panel of Fig.~\ref{fg-b2}.

Comparison of results of computations for two models, allows to make the
following conclusions. Magnetic field of the stellar wind is an important
factor, affecting the process of the outflow of the ionospheric envelope from
the Roche lobe of a hot Jupiter. In the case of a weak wind field, the main
limiting factor is dynamic wind pressure. As the field increases, the total wind
pressure increases. As a result, for the same other parameters, the dimensions
of the quasi-closed ionospheric envelope decrease. In the model 2 in the direction
of the star ($x$-axis) the size of the envelope turned out to be by a factor 1.5
smaller compared to the case of weak field (model 1). In model 1, in the
direction of the orbital motion of the planet ($y-$ axis) the envelope moves
away from the planet to the distance of about 10 photometric radii, whereas in
model 2 this distance is approximately 5 photometric radii. Note, just these
characteristics of the envelope (in the direction of the orbital motion of the
planet) determine the observed phenomena during the transit, associated
with the early onset of the eclipse in the near ultraviolet range \cite
{Vidal2003}. Consequently, the observed properties of the early onset of eclipse
during the transit are also dependent on the strength of the wind magnetic
field. 

\section{Conclusion}

The analysis performed in this paper leads to the conclusion that many hot
Jupiters may be located in the sub-Alf\'{v}en zone of the stellar wind of the
parent star. This means that in the studies of the flow of the stellar wind
around the atmosphere of a hot Jupiter magnetic field of the wind is an
extremely important factor, consideration of which is absolutely necessary, both
in theoretical models and in the interpretation of observational data. The fact
is that in the sub-Alf\'{v}en zone magnetic pressure of the stellar wind exceeds
its dynamic pressure even if the orbital motion of the planet is taken into
account.

Based of rather simple model considerations, as well as summarizing the results
of numerical experiments, we suggested a classification of possible shapes of
magnetospheres of hot Jupiters. In particular, well studied magnetospheres of the
Earth and Jupiter in our classification belong to the subtype \textit{A1}
(intrinsic magnetosphere
with bow shock) with closed envelopes. As it was shown by the analysis of
observational data, the magnetospheres of many hot Jupiters can belong to the
subtype \textit{B2} (shock less induced
magnetosphere). In this case, magnetic
field of the wind is rather strong and, therefore, the flow of the stellar wind
around the atmosphere of the planet appears to be shock-less. Detached shock
waves around the atmosphere and the outflow from the Lagrange point $L_1$ do not
form. The structure of such a magnetosphere is fundamentally different from the
magnetosphere of terrestrial type.

However, since the characteristics of the stellar wind can vary quite a bit over
time (approximately, by a factor 1.5 to 2), a fraction of hot Jupiters falls
into the parameter space, which we figuratively named ``gray area''. In this
zone, the type of stellar wind flow around the planet is intermediate between 
a shock
and a shock-less flow. The study of the structure of the magnetospheres of this
type is a separate task.

To study the process of the flow of stellar wind around hot Jupiters with
simultaneous consideration of both the planet magnetic field and the wind
magnetic field, we developed a relevant three-dimensional magnetohydrodynamical
numerical model. The basis of our numerical model is Roe-Einfeldt-Osher
difference scheme of higher order approximation for the equation of ideal
magnetohydrodynamics. In our numerical model, the total magnetic field is
represented as a superposition of the external magnetic field and the magnetic
field induced by electric currents in the plasma itself. As an external field we
applied superposition of the dipole magnetic field of the planet and the radial
component of the wind magnetic field. In the numerical algorithm, the factors
associated with the presence of the external magnetic field were taken into
account at a separate step using appropriate difference scheme of Godunov type.

We calculated two models that differ by the strength of the average magnetic
field at the surface of the star only. In the first model the wind magnetic
field was weak and the flow pattern matched well both purely gas-dynamical
calculations \cite{Bisikalo2013} and calculations taking into account the
magnetic field of the planet only \cite{Arakcheev2017}. These models give a
similar picture of the supersonic flow around the planet, since the proper
magnetic field of a hot Jupiter is quite weak. In the terms of our
classification, the corresponding magnetosphere belongs to the \textit{A1}
subtype (intrinsic magnetosphere
with bow shock) with an open ionospheric envelope. For
such parameters the planet is in the super-Alf\'{v}en zone of the wind and in
its interaction with the wind a detached shock wave is formed. In the second
model, the magnetic field of stellar wind corresponded to the magnetic field of
the solar wind, which is defined by the average magnetic field of the quiet Sun.
In this case, the hot Jupiter falls into the sub-Alf\'{v}en wind zone and,
therefore, the detached shock wave does not form, just as it observed in the
calculations. In terms of our classification, such a magnetosphere belongs to
the subtype \textit{B2} (shockless induced magnetosphere) with an open
ionospheric envelope.

\section*{Acknowledgements}

The authors acknowledge P.V. Kaigorodov for useful discussions. 
This study was supported by RSF (project 18-12-00447). Computations were
carried out using the supercomputer of the National Research Center ``Kurchatov Institute''.

\end{document}